\documentclass[fleqn,usenatbib]{mnras}
\usepackage{newtxtext,newtxmath}
\usepackage[T1]{fontenc}
\usepackage{ae,aecompl}
\usepackage{epsfig,amsmath,amsfonts,graphicx,subfigure,enumitem,url, deluxetable}
\usepackage{epstopdf}
\usepackage{hyperref}
\usepackage{multirow}
\usepackage{booktabs}
\usepackage{times}
\usepackage{caption}
\usepackage{newtxtext,newtxmath}
\usepackage{{threeparttable}}
\hypersetup{linkcolor=blue,citecolor=blue,filecolor=cyan,urlcolor=magenta}

\title[A disc-dominated IP IGR~J15094-6649]{Unravelling optical and X-ray properties of the disc-dominated intermediate polar IGR~J15094-6649}
\author[Joshi et al.]{
Arti Joshi$^{1}$\thanks{E-mail: arti.joshi@iiap.res.in},
Nikita Rawat$^{2}$,
Axel Schwope$^{3}$,
J. C. Pandey$^{2}$,
Simone Scaringi$^{4}$,
D. K. Sahu$^{1}$, 
Srinivas M Rao$^{2}$,
\newauthor ~and Mridweeka Singh$^{1}$
\\~\\
$^{1}$Indian Institute of Astrophysics, Koramangala, Bangalore 560 034, India\\
$^{2}$Aryabhatta Research Institute of observational sciencES, Manora Peak, Nainital 263 001, India\\
$^{3}$Leibniz-Institut für Astrophysik Potsdam (AIP), An der Sternwarte 16, 14482 Potsdam, Germany\\
$^{4}$Centre for Extragalactic Astronomy, Department of Physics, Durham University, South Road, Durham DH1 3LE, UK
}

\date{Accepted XXX. Received YYY; in original form ZZZ}
\pubyear{2022}
\begin{document}
\label{firstpage}
\pagerange{\pageref{firstpage}--\pageref{lastpage}}
\maketitle

\begin{abstract}
We present analyses of an Intermediate Polar, IGR~J15094-6649, based on the archival optical data obtained from the Transiting Exoplanet Survey Satellite (\textit {TESS}) and X-ray data obtained from the \textit{Suzaku}, \textit{NuSTAR}, and Neil Gehrels Swift Observatory (\textit{Swift}). Present analysis confirms and refines the previously reported spin period of IGR~J15094-6649 as 809.49584$\pm$0.00075 s. Clear evidence of a beat period of 841.67376$\pm$0.00082 s is found during the long-term \textit{TESS} optical observations, which was not evident in the earlier studies. The dominance of X-ray and optical spin pulse unveils the disc-fed dominance accretion, however, the presence of an additional beat frequency indicates that part of the accreting material also flows along the magnetic field lines. The energy-dependent spin pulsations in the low ($<$10 keV) energy band are due to the photoelectric absorption in the accretion flow. However, the complex absorbers may be responsible to produce low amplitude spin modulations via Compton scattering in the hard ($>$10 keV) energy band and indicate that the height of the X-ray emitting region may be negligible. The observed double-humped X-ray profiles with a pronounced dip are indicative of the photoelectric absorption in the intervening accretion stream. Analysis of the X-ray spectra reveals the complexity of the X-ray emission, being composed of multi-temperature plasma components with a soft excess, reflection, and suffers from strong absorption. 
\end{abstract}


\section{Introduction}
\label{sec:intro}
\begin{table*}
\small
\caption{Log of the \textit{TESS}, \textit{Suzaku}, \textit{NuSTAR}, and \textit{Swift}  
 observations of IGR1509.\label{tab:obslog}}
\setlength{\tabcolsep}{0.05in}
\centering
\begin{tabular}{@{}cccccccc@{}}
\hline
Satellite  &Instrument   &Sector or       & Date of Obs.  &Time of Obs.  & Integration Time             & Mean Flux (e$^-$/s)   \\
	    &            &Observation ID    &(YYYY-MM-DD)   &(UT)          & (ks)      	         &  or   Mean Count Rate (Cts/s)        \\
\hline 
\textit{TESS}      & Photometer  &Sector 12      & 2019-05-21   & 10:45:32    & 2413.6        & 169.5$\pm$0.1 \\
	            & Photometer  &Sector 38      & 2021-04-29   & 08:34:07    & 2306.5        & 202.4$\pm$0.1\\
	            & Photometer  &Sector 39      & 2021-05-27   & 06:34:49    & 2414.5        & 214.2$\pm$0.1\\
\textit{Suzaku}    & XIS-FI      &405007010      & 2011-01-27   & ~16:52:31   & 49.5          & 1.031$\pm$0.004 \\  
		  & XIS-BI      &405007010      & 2011-01-27   & ~16:52:31   & 49.5          & 0.531$\pm$0.003 \\  
		  & HXD-PIN     &405007010      & 2011-01-27   & ~16:52:31   & 48.0          & 0.085$\pm$0.004 \\  
\textit{NuSTAR}    & FPMA/FPMB   &30460013002    & 2018-07-19   & ~23:24:37   & 41.3          & 1.74$\pm$0.01\\        
\textit{Swift}    & XRT         &00088619001    & 2018-06-23   & ~17:05:42   & 0.3           & 0.28$\pm$0.03\\
           	  & XRT         &00088619002    & 2018-07-19   & ~23:45:03   & 0.7           & 0.20$\pm$0.02\\
	          & XRT         &00088619003    & 2018-07-20   & ~00:59:57   & 6.0           & 0.20$\pm$0.01\\
\hline
\end{tabular}
\end{table*}

IGR~J15094-6649 (hereafter IGR1509) is an intermediate polar (IP) in which a white dwarf (WD) accretes material from the mass donating secondary via Roche-lobe overflow \citep{Warner95}. The IPs are known as asynchronized binaries (\textit {i.e.,} $P_\omega$$<$$P_\Omega$, where $P_\omega$ and $P_\Omega$ are spin and orbital periods, respectively) with a WD magnetic field strength of $<$10 MG. The optical and X-ray signals in IPs are modulated on the spin period, orbital period, beat period ($P_\omega$$_-$$_\Omega$), and other side-bands \citep{Patterson94}. A characteristic feature of these systems is the presence of multiple periodicities in the X-ray and optical bands due to complex interactions between the spin and orbital modulations either through variations of the accretion rate or through the reprocessing of primary radiation by other components of the system. The  existence of multiple periodic components is an important diagnostic tool for determining the IP nature of a potential member of this class. Moreover, the accretion occurs in these systems with three different mechanisms, viz; disc-fed, disc-less, and disc-overflow \citep[see][]{Hameury86, Rosen88, Lubow89, Ferrario99}. An accretion disc is usually formed in the IPs and truncated at the magnetospheric radius where the magnetic pressure exceeds the ram pressure and subsequently the accreting material channels in accretion columns towards the poles of the WD and referred to as `disc-fed' IPs \citep{Rosen88}. In the `disc-less (or stream-fed)' IPs, the accreting material is directly channeled toward the magnetic pole of the WD without an intervening disc with the high magnetic fields of the WD \citep{Hameury86, Hellier91, Wynn92}. A combination of both disc-fed and stream-fed \citep[known as `disc-overflow';][]{Hellier95} accretion also occurs in the IPs, where an accretion disc is present, but matter from the stream overpasses the disc \citep{Hellier89, Lubow89, Armitage96}. These modes of accretions can be identified with the presence of a wide range of X-ray and optical frequencies \citep{Wynn92, Norton96}. In the disc-fed accretion, strong modulations are expected to occur at the WD spin frequency \citep{Kim95}. However, in the pure stream-fed systems, the beat frequency is dominating \citep{Wynn92}. For a disc-overflow accretion, modulations at both spin and beat frequencies are expected to occur and the main difference lies in the varying power/amplitude between the two \citep{Hellier93, Norton97}. 

In the IPs, the magnetically channeled accretion column impacts the WD surface, and strong shocks are formed in the accretion columns that heat the plasma up to a high temperature, about $\sim$100 MK. The shocked gas subsequently cools as it falls towards the surface of the WD via thermal bremsstrahlung emitting hard X-rays \citep{Aizu73}. So, a multi-temperature plasma model is generally required to quantify the temperature distribution in the post-shock region (PSR) \citep{Done95}. The UV and soft X-ray components ($<$2 keV) mainly arise because of the reprocessing of hard X-ray photons from the surface of the WD, however, the cyclotron radiation at optical wavelength is seen due to the reprocessing of high energy X-rays in the surface layers of the disc (including the hotspot) and/or the atmosphere of the secondary. The soft X-ray spectral component can be modeled using blackbody emission which is usually dominant in polars. However, some IPs also possess soft X-ray emissions which can be modeled by using a blackbody component with a temperature more than that of the polars and are referred to as `soft-IPs' \citep[for detail see,][]{Haberl95, Evans07}. The observed radiation in IPs interacts with its surroundings like the WD surface, the accretion disc, the accretion curtain, and any circumstellar medium that might exist and undergo photoelectric absorption which produces strong modulations and these modulations are more pronounced in the soft energy bands. The high-energy X-rays are also expected to be highly absorbed by cold matter and reprocessed and/or reflected by the WD surface \citep{Matt91, Done92, Beardmore95}. The detection of the reflection in the IPs can be confirmed with the observed Compton reflection hump and the presence of fluorescent Fe K$\alpha$ emission line in their X-ray spectra. Therefore, a broadband X-ray spectroscopy is required to understand the effect of both absorption and reflection.

IGR1509 was discovered by \cite{Revnivtsev06} in the INTEGRAL/IBIS all-sky survey in the 17-60 keV energy range. \cite{Barlow06} further explored the INTEGRAL/IBIS data in the 20-100 keV energy band and derived a best-fitted temperature of 13.8$\pm$5.1 keV along with the X-ray flux of 1.38$\times$10$^{-11}$ erg s$^{-1}$ cm$^{-2}$. Based on the optical spectroscopy, \citet{Masetti06} tentatively classified IGR1509 as an IP. Later, \citet{Pretorius09} provided a clear detection of the orbital and spin periods of 5.89$\pm$0.01 h and 809.424$\pm$0.018 s using spectroscopic and photometric observations, respectively, and classified this system as an IP. \cite{Butters09} further confirmed the IP classification of IGR1509 with the detection of the X-ray spin period of 809.7$\pm$0.6 s using the \textit{RXTE} observations. Strong spin-modulated circular polarization was detected for IGR1509 \cite[see][]{Potter12}, consistent with cyclotron emission from a WD. Based on the detection of strong circular polarization, it was suggested that along with the disc, IGR1509 has an extended accretion curtain and may have a connection with either polars or soft X-ray-emitting IPs. Later, \cite{Bernardini12} reported the X-ray spin period of 808.7$\pm$0.1 s along with the harmonic of the beat and  $2\omega$-$\Omega$ side-band using the \textit{XMM-Newton} observations and interpreted that IGR1509 belongs to the class of disc-overflow IPs. Recently, \cite{Shaw20} presented a legacy survey of 19 MCVs, including IGR1509 with the \textit{NuSTAR}. They fitted the \textit{NuSTAR} spectra in the 20-78 keV energy band using the PSR X-ray spectral model and derived the WD mass of 0.73$\pm$0.06 $M_\odot$ for IGR1509. In the light of earlier studies, the beat period was not clearly detected which is essential to probe the true nature or accretion geometry of an IP. Therefore, with a motivation to ascertain its true accretion geometry, we present the detailed optical analysis using the long, uninterrupted, high-cadence \textit{TESS} observations. Moreover, the broad-band \textit{Suzaku} and contemporaneous \textit{Swift} and \textit{NuSTAR} X-ray observations encouraged us to unveil the nature of this IP in much greater detail. This paper is organized as follows: Section \ref{sec:obs} summarizes archival optical and X-ray 
 observations and their data reduction description. Analyses and the results of the optical and X-ray data are described in Section \ref{sec:analysis}. Finally, we present a discussion and conclusions in Sections \ref{sec:diss} and \ref{sec:conc}, respectively.

\begin{figure*}
\centering
\subfigure[]{\includegraphics[width=0.8\textwidth]{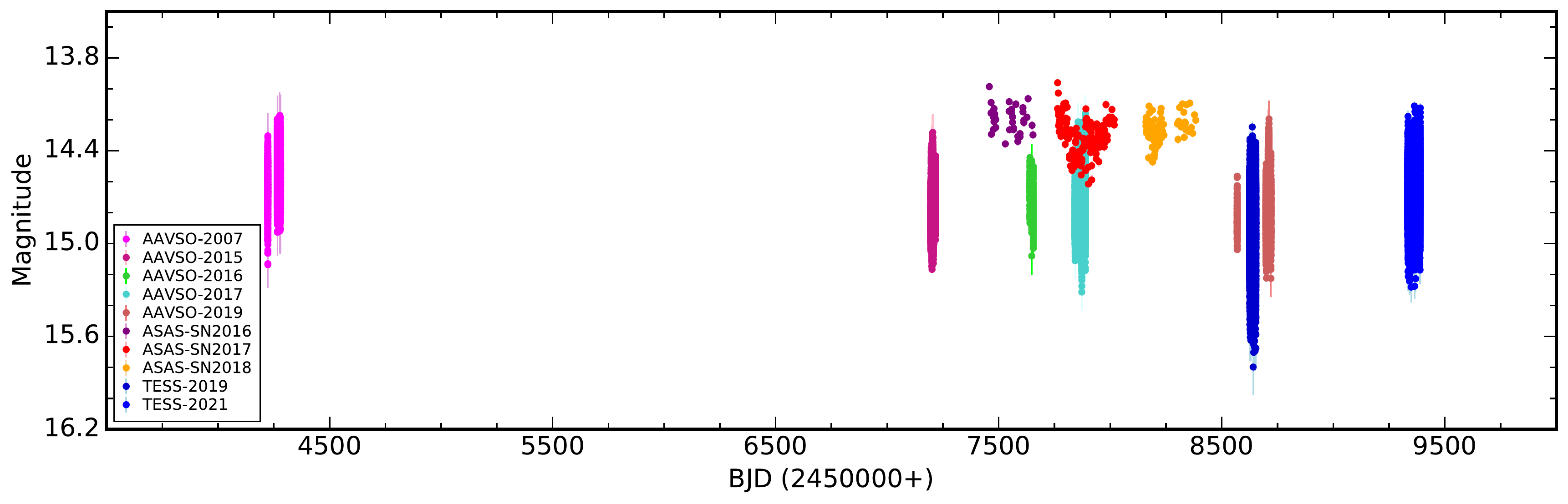}\label{fig:longtermlc}}
\subfigure[]{\includegraphics[width=0.8\textwidth]{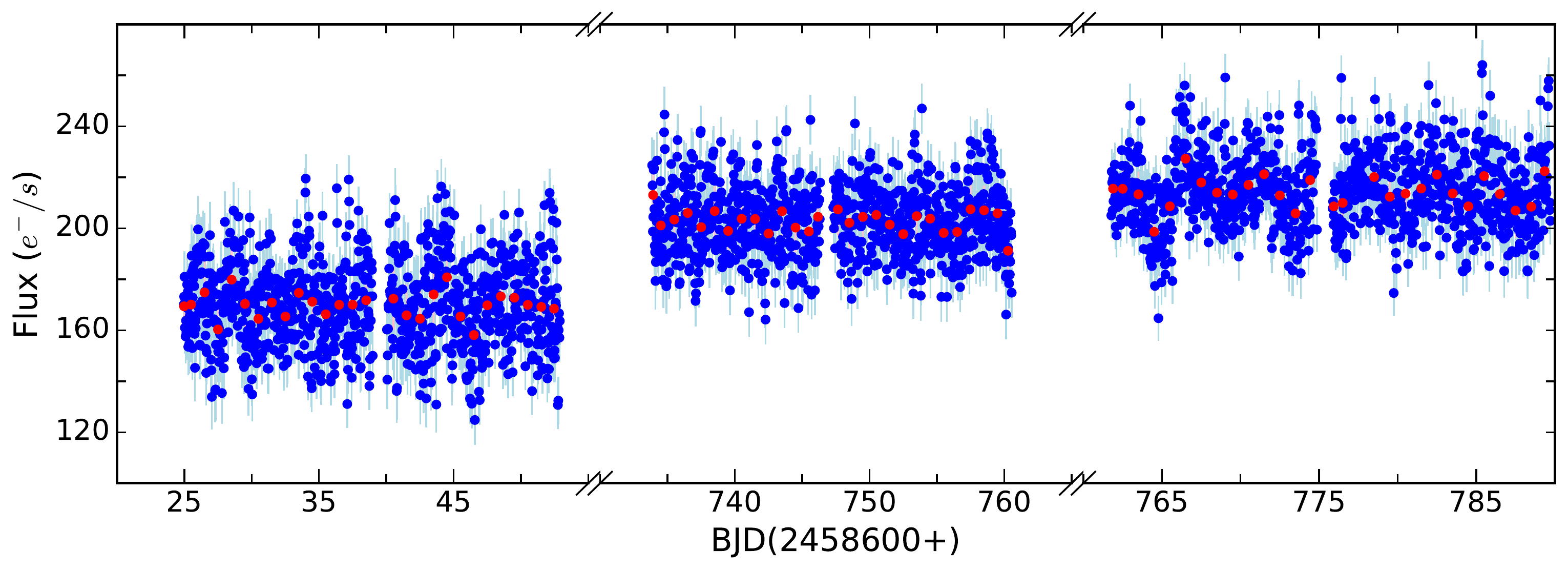}\label{fig:tessfulllc}}
\subfigure[]{\includegraphics[width=0.8\textwidth]{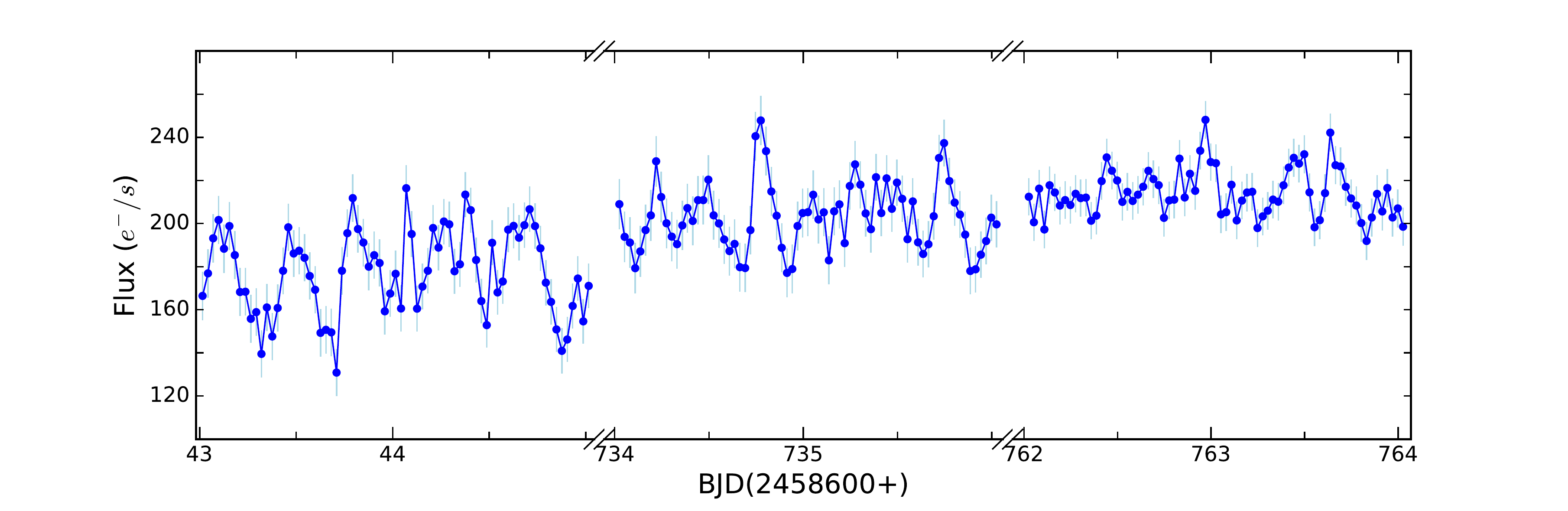}\label{fig:tesspartlc}}
\caption{(a) Long-term light curve of IGR1509 as observed from AAVSO, ASAS-SN, and \textit{TESS}, (b) the full \textit{TESS} light curves of IGR1509 for sectors 12, 38, and 39, respectively, where red dots represent the mean flux of each day, and (c) the zoomed version of the two consecutive days of \textit{TESS} observations for sectors 12, 38, and 39, respectively.}
\end{figure*}


\section{Observations and Data Reduction}\label{sec:obs}
A detailed log of optical and X-ray observations of IGR1509 is given in Table \ref{tab:obslog}. We have described the details of each observation in the forthcoming subsections.

\subsection{Optical observations} \label{sec:optobs}
IGR1509 was observed by \textit{TESS} with camera 2 in sectors 12, 38, and 39 on 21 May 2019 at UT 10:45:32, 29 April 2021 at UT 08:34:07, and 27 May 2021 at UT 06:34:49, respectively at a cadence of 2 minutes. There is a gap of almost 22 months 9 days between the observations in sectors 12 and 38, however, a gap of 1 day 5 hrs between sectors 38 and 39. The \textit{TESS} bandpass extends from 600 to 1000 nm with an effective wavelength of 800 nm \citep[see][for details]{Ricker15}. \textit{TESS} observations are broken up into sectors, each lasting two orbits, or about 27.4 days, and conducts its downlink of data while at perigee. This results in a small gap in the data compared to the overall run length. The data were stored under Mikulski Archive for Space Telescopes (MAST) data archive \footnote{\textcolor{magenta}{\url{https://mast.stsci.edu/portal/Mashup/Clients/Mast/Portal.html}}} with identification number `TIC 261698173'. The \textit{TESS} pipeline provides two flux values: simple aperture photometry (SAP) and pre-search data conditioned SAP (PDCSAP). The PDCSAP light curve attempts to remove instrumental systematic variations by fitting and removing those signals that are common to all stars on the same CCD \footnote{\textcolor{black}{see section 2.1 of \textit{TESS} archive manual available at \textcolor{magenta}{\url {https://outerspace.stsci.edu/display/TESS/2.0+-+Data+Product+Overview}}}}. We saw no significant differences in the two light curves and elected to use the PDCSAP data for our analysis. Data taken during an anomalous event had quality flags greater than 0 in the FITS file. We have considered only the data points with the `QUALITY flag' = 0.

We also utilized the public photometry database of the All-Sky Automated Survey for Supernovae \citep[ASAS-SN\footnote{\textcolor{magenta}{\url{https://asas-sn.osu.edu/variables}}};][]{Shappee14, Kochanek17} and downloaded long-term light curves for IGR1509. Further, the optical photometry data of IGR1509 was also collected from the American Association of Variable Star Observers \citep[AAVSO\footnote{\textcolor{magenta}{\url{https://www.aavso.org/}}};][]{Kafka21} database, which was observed in between 2007 to 2019. Its long-term AAVSO and ASAS-SN light curves are easily accessible online.

\begin{figure}
\centering
\includegraphics[width=\columnwidth]{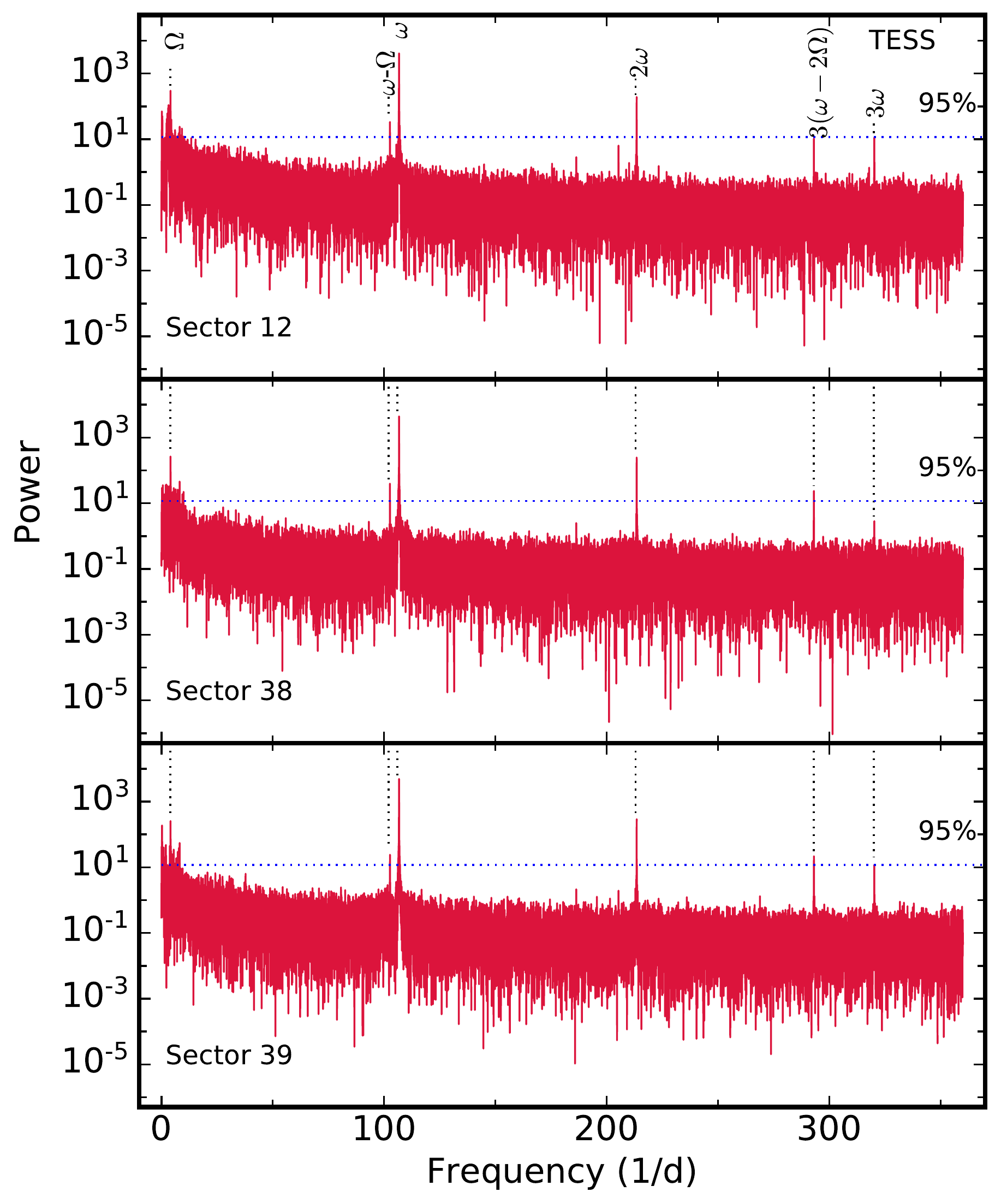}
\caption{LS power spectra as obtained from the \textit{TESS} observations for sectors 12, 38, and 39.}
\label{fig:tessscps}
\end{figure}

\begin{figure}
\centering
\includegraphics[width=\columnwidth]{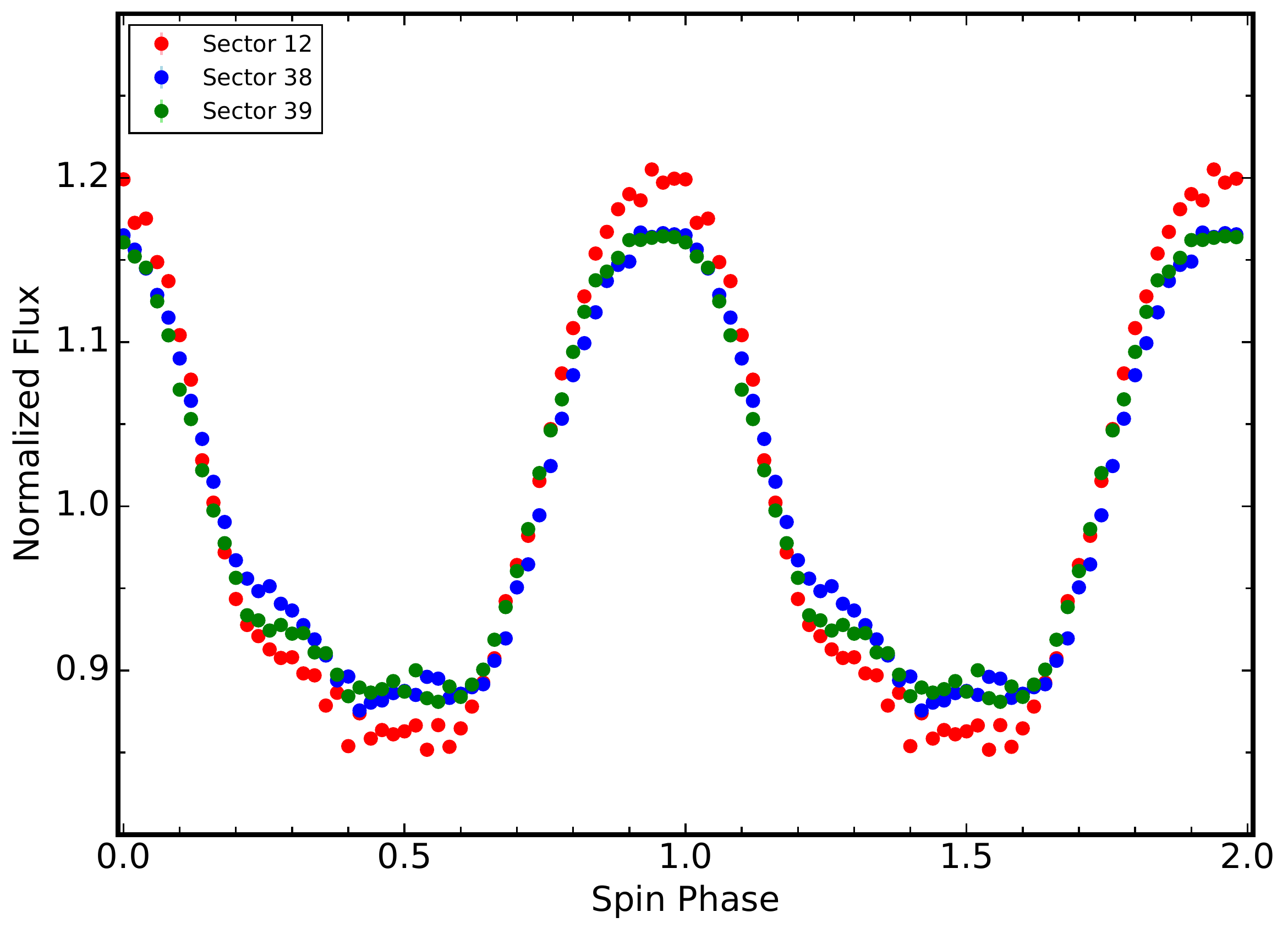}
\caption{Spin pulse profiles with phase bin of 0.02 as obtained from the \textit{TESS} observations for sectors 12, 38, and 39.}
\label{fig:tessspinflc}
\end{figure}

\subsection{X-ray observations} \label{sec:xrayobs}
\subsubsection{\textit{Suzaku}}
The \textit{Suzaku} observations of IGR1509 were carried out on 27 January 2011 at 16:52:31 (UT) with an offset of 0.366 arcmin, using X-ray Imaging Spectrometers \citep[XIS;][]{Koyama07} and Hard X-ray Detector \citep[HXD;][]{Takahashi07}. XIS and HXD observe in the 0.2-12.0 keV and 10-600 keV energy ranges, respectively. The HXD consists of two detectors: the GSO well-type phoswich counters and the silicon PIN diodes. For XIS and HXD-PIN detectors the exposure times were 49.5 ks and 48.0 ks, respectively. Using \textsc{xselect}, the task \textsc{aepipeline} (version 1.1.0) with the latest calibration files was used for reprocessing the XIS data. The data was reduced under the standard screening criteria. The task \textsc{aebarycen} was used for the barycentric corrections in the event files. The X-ray light curves and spectra from XIS instruments were extracted from a circular region with a radius of 180$\arcsec$ around the source whereas the background was taken near the source-free region with the same radius as the source region. The rmf and arf files were generated for the XIS detector using the FTOOLS tasks \textsc{xisrmfgen} and \textsc{xissimarfgen}. The background-subtracted light curves and spectra were generated from a back-illuminated (BI-XIS1) CCD camera and two front-illuminated (FI) CCD cameras (XIS0 and XIS3). XIS0 and XIS3 light curves were combined using \textsc{lcmath} \citep{Blackburn95} task for further timing analysis. Spectra from FI CCDs were added using \textsc{addascaspec}. The FI and BI spectra were grouped using \textsc{grppha} to have at least 20 counts per bin. 

To analyze the HXD-PIN data, we have used the non-X-ray background (NXB) files and response matrix files provided by the HXD team\footnote{\textcolor{magenta}{\url{http://www.astro.isas.jaxa.jp/suzaku/analysis/hxd/}}}. We have downloaded the appropriate `tuned' background (updated version of `$bgd\_d$'; METHOD=LCFITDT and version of METHODV=`2.0ver0804') files for our observation. Since the background event files have own good time intervals (GTIs), so to merge the good time intervals and to obtain a common value for the PIN background and source event files, we have used the task \textsc{mgtime}. We then extracted the source and background spectra with GTI. We have corrected HXD-PIN data for the dead time of the observed spectrum, as required. The exposure of the PIN background was also corrected to compensate 10 times higher event rate of PIN background event file than the real background. Finally, the contribution from the cosmic X-ray background was subtracted from the source spectrum, using the appropriate cosmic X-ray background flat response file provided by the HXD team.

\begin{table*}
\small
\centering
\caption{Periods corresponding to dominant peaks in the power spectra of IGR1509 obtained from the LS periodogram analysis of the \textit{TESS}, \textit{Suzaku} (0.3-10.0 keV), \textit{NuSTAR} (3-78 keV), and \textit{Swift} (0.3-10.0 keV) observations.}\label{tab:ps}
\setlength{\tabcolsep}{0.035in}
\begin{tabular}{lccccccccccc}
\hline
Telescopes   & Epoch & \multicolumn{6}{c}{Periods} \\
\cline{3-8}\\
	                       &                 &  $P_{\Omega}$  & $P_($$_\omega$$_-$$_{\Omega}$$_)$          & $P_{\omega}$ &  P$_{2\omega}$ & P$_3$$_($$_{\omega}$$_-$$_{2\Omega}$$_)$ & P$_{3\omega}$  \\
	     &           & ( h)         & ( s )                                      & ( s )        & (s)             & ( s )                                   & ( s )            \\
       
\hline\\
\textit{TESS}     	   & 2019-05-21           & $5.86\pm0.01$      &$841.73\pm0.07$  &$809.47\pm0.06$    &$404.74\pm0.02$  &  $294.80\pm0.01$  & $269.83\pm0.01$ \\
          	   & 2021-04-29           & $5.87\pm0.01$       &$841.73\pm0.07$  &$809.46\pm0.07$    &$404.74\pm0.02$  &  $294.83\pm0.01$  & $269.82\pm0.01$ \\
          	   & 2021-05-27           & $5.87\pm0.01$        &$841.65\pm0.07$  &$809.48\pm0.06$    &$404.74\pm0.02$  &  $294.80\pm0.01$  & $269.83\pm0.01$ \\
                   & Combined$^\dagger$   & $5.87213\pm0.00014$ &$841.67376\pm0.00082$  &$809.49584\pm0.00075$ &$404.74791\pm0.00018$  &  $294.81353\pm0.00010$  & $269.82857\pm0.00008$ \\
\textit{Suzaku}       & 2011-01-27           & ....	        &....             &$809.14\pm2.22$    &$404.57\pm0.55$  &  ....             & $269.71\pm0.25$\\
\textit{NuSTAR}       & 2018-07-19           & ....	        &$840.57\pm2.40$  &$808.15\pm2.23$    &$405.19\pm0.55$  &  ....             & $270.12\pm0.25$ \\
\textit{Swift}       & 2018-07-19/20        & ....	        &$834.39\pm5.08$  &$809.70\pm4.79$    &$404.85\pm1.19$  &  ....             & $269.90\pm0.53$ \\
\hline
\end{tabular}
\tablecomments{$\dagger$ represents the periods derived from the combined 
 \textit{TESS} observations of all three sectors 12, 38, and 39.} 
\end{table*}

\begin{figure*}
\centering
\subfigure[Sector 12]{\includegraphics[width=0.82\textwidth]{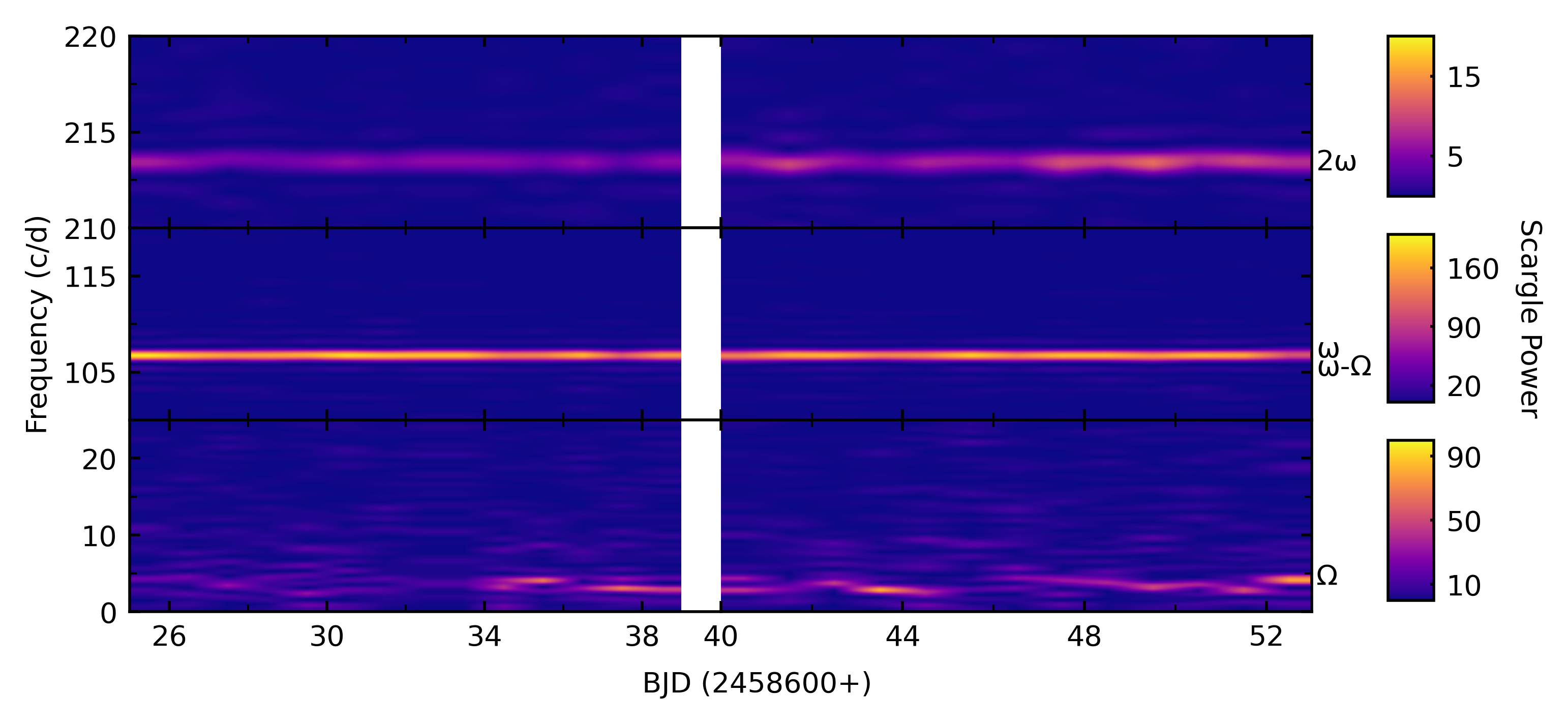}\label{fig:tess1daypssec12}}
\vskip -6pt
\subfigure[Sector 38]{\includegraphics[width=0.82\textwidth]{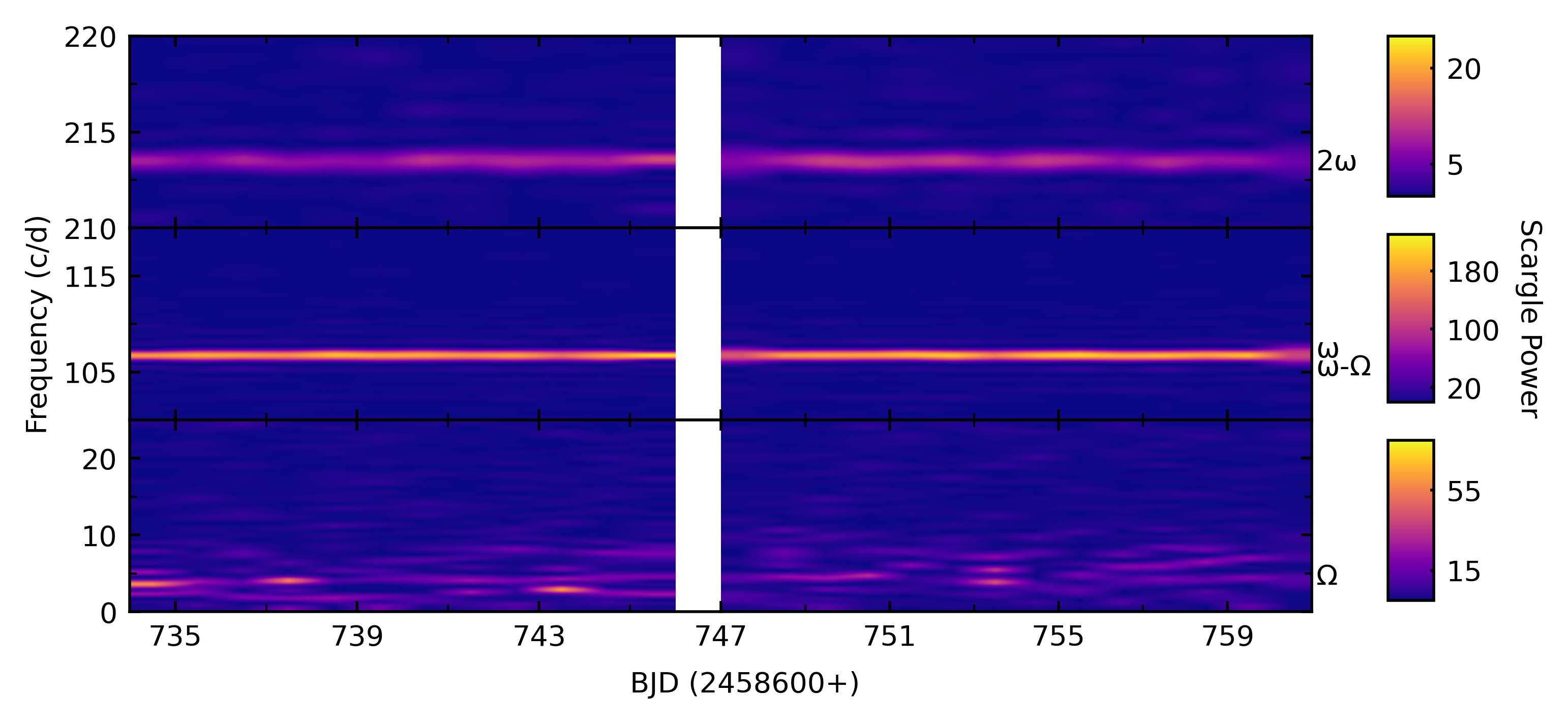}\label{fig:tess1daypssec38}}
\vskip -6pt
\subfigure[Sector 39]{\includegraphics[width=0.82\textwidth]{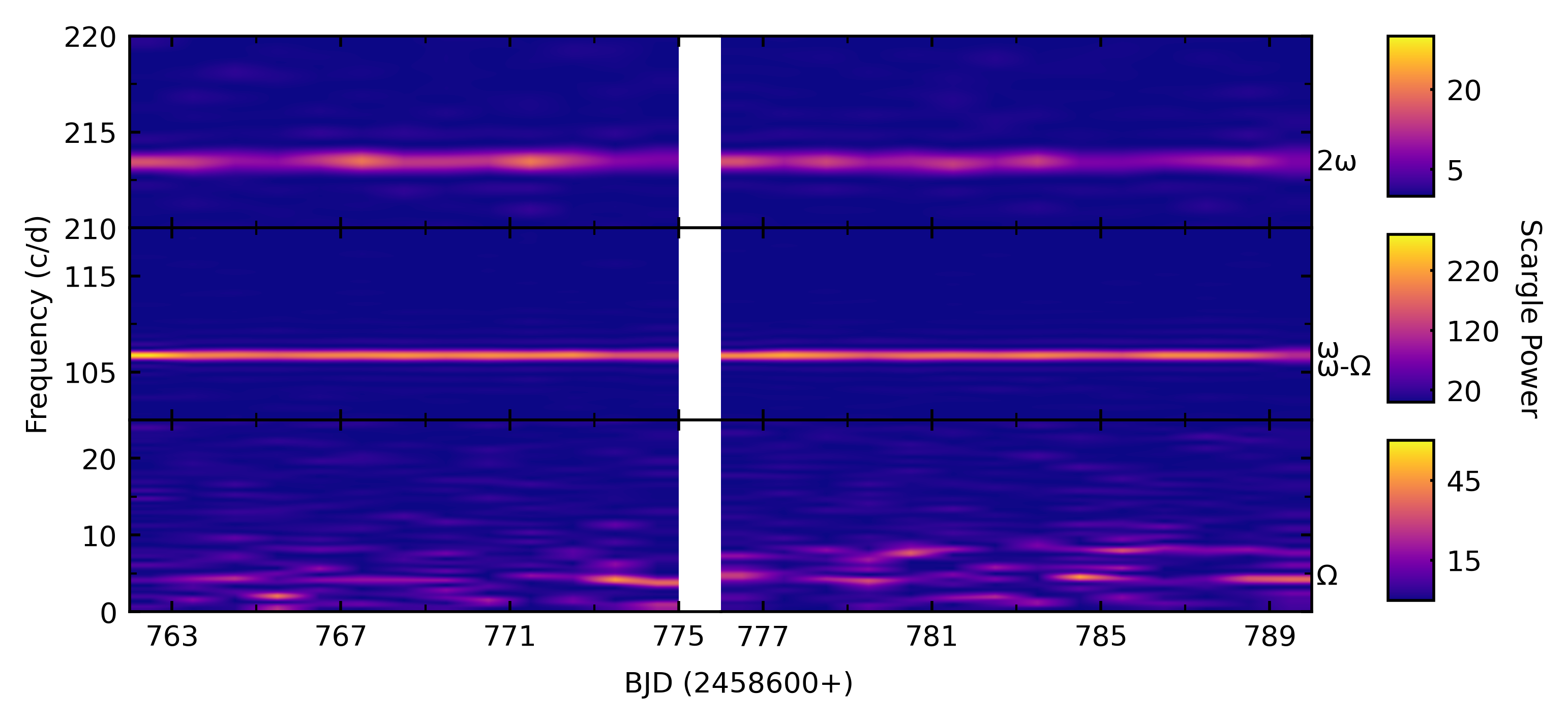}\label{fig:tess1daypssec39}}
\vskip -6pt
\caption{One-day time-resolved power spectra as obtained from the 
\textit{TESS} observations for sectors 12, 38, and 39 in the $\Omega$, $\omega-\Omega$, $\omega$, and $2\omega$ frequency region.}
\end{figure*}

\begin{figure*}
\centering
\subfigure[Sector 12]{\includegraphics[width=58mm, height=50mm]{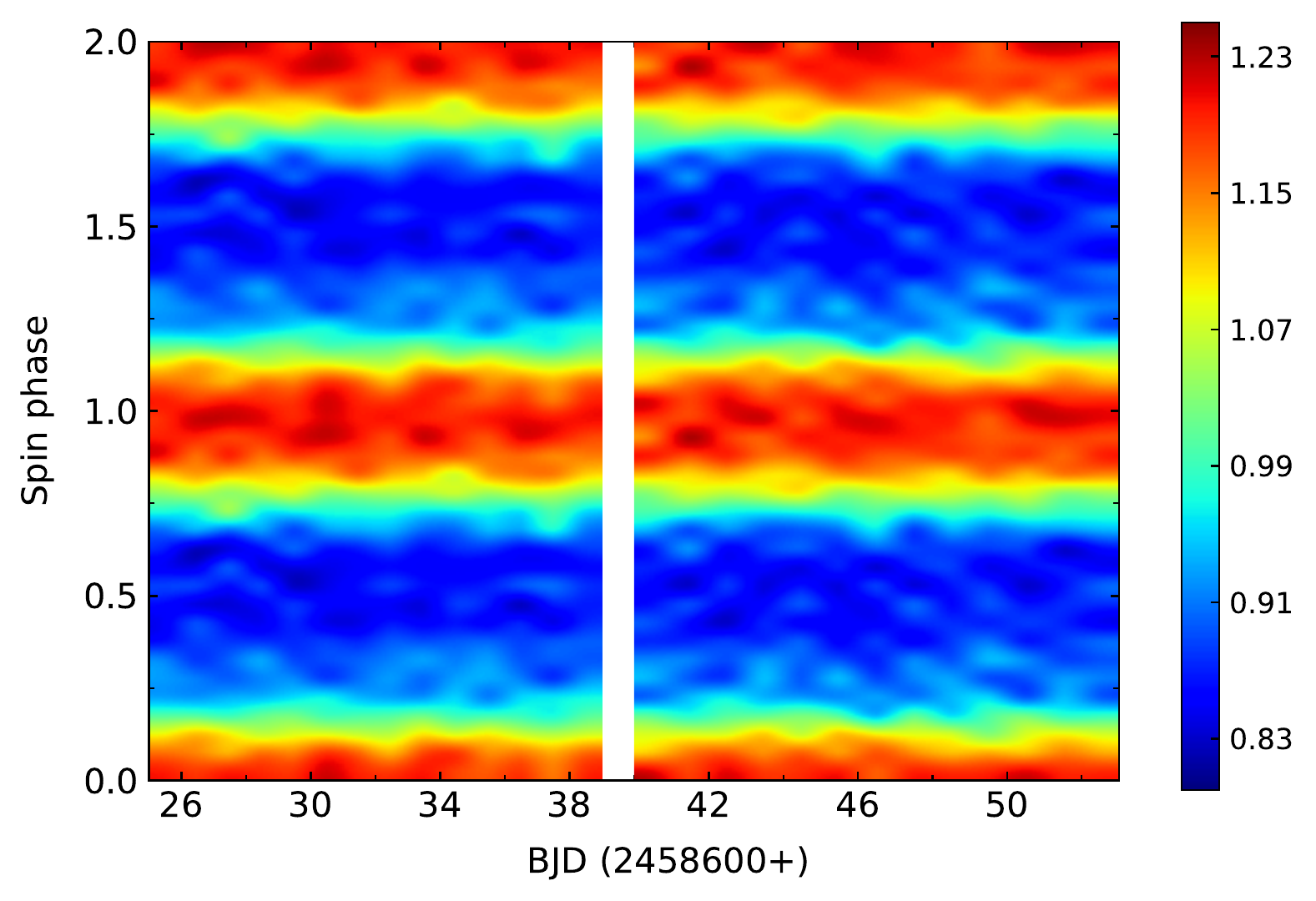}}
\subfigure[Sector 38]{\includegraphics[width=58mm, height=50mm]{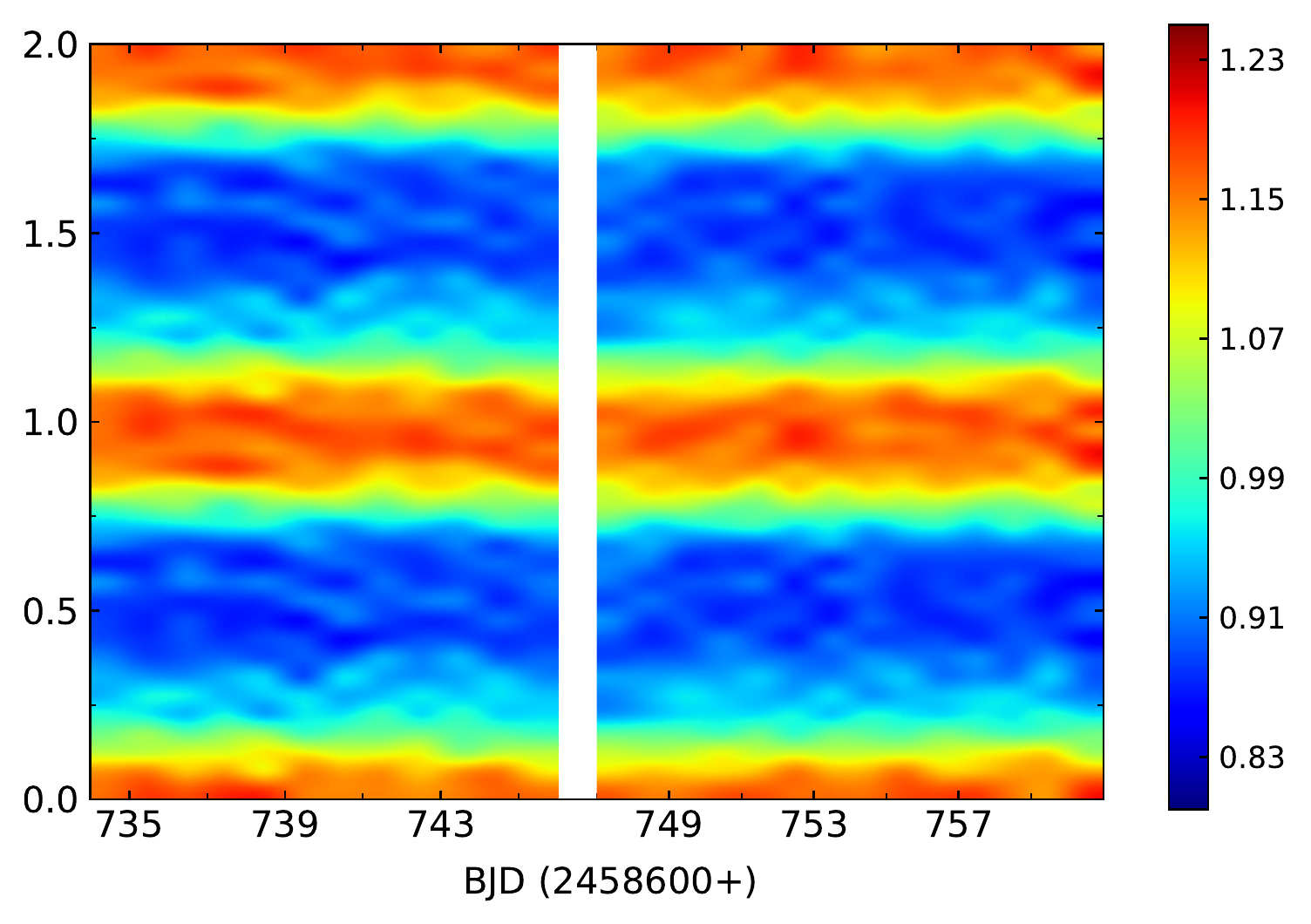}}
\subfigure[Sector 39]{\includegraphics[width=58mm, height=50mm]{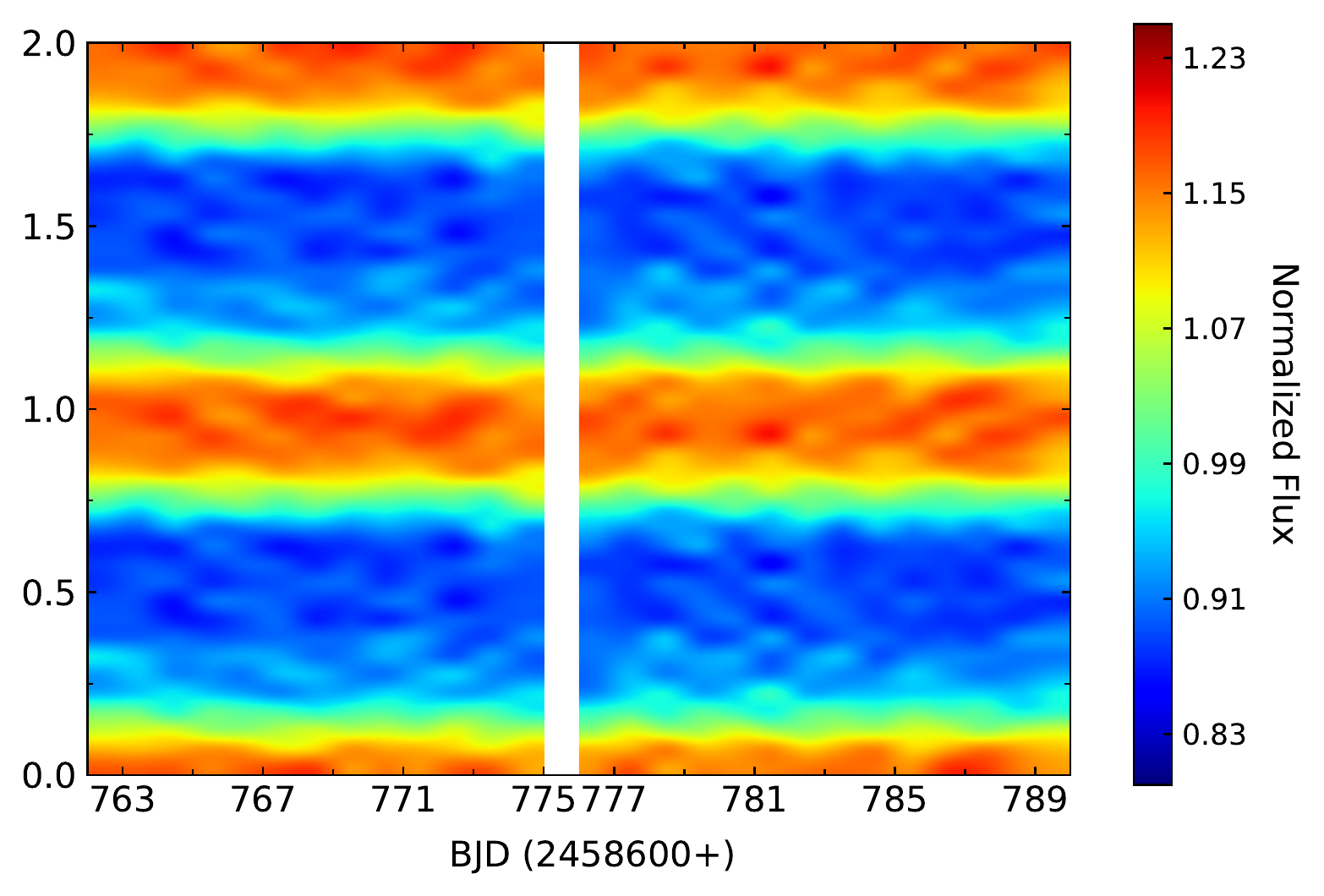}}
\caption{Evolution of the spin pulse profiles during the \textit{TESS} observations for sectors 12, 38, and 39.}
\label{fig:tess1dayflc}
\end{figure*}

\begin{figure}
\centering
\subfigure[]{\includegraphics[width=85mm, height=85mm]{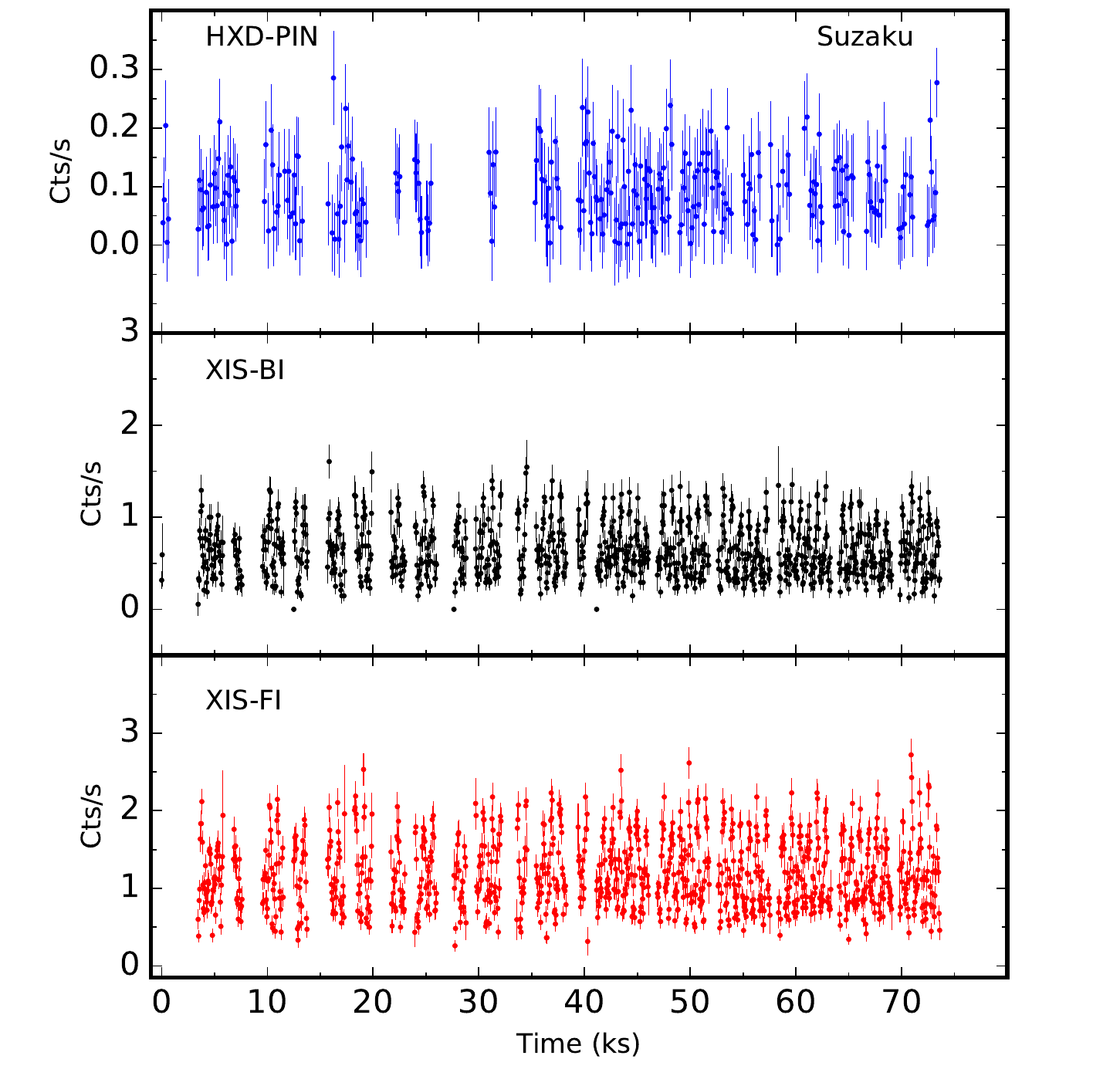}\label{fig:szlc}}
\subfigure[]{\includegraphics[width=\columnwidth]{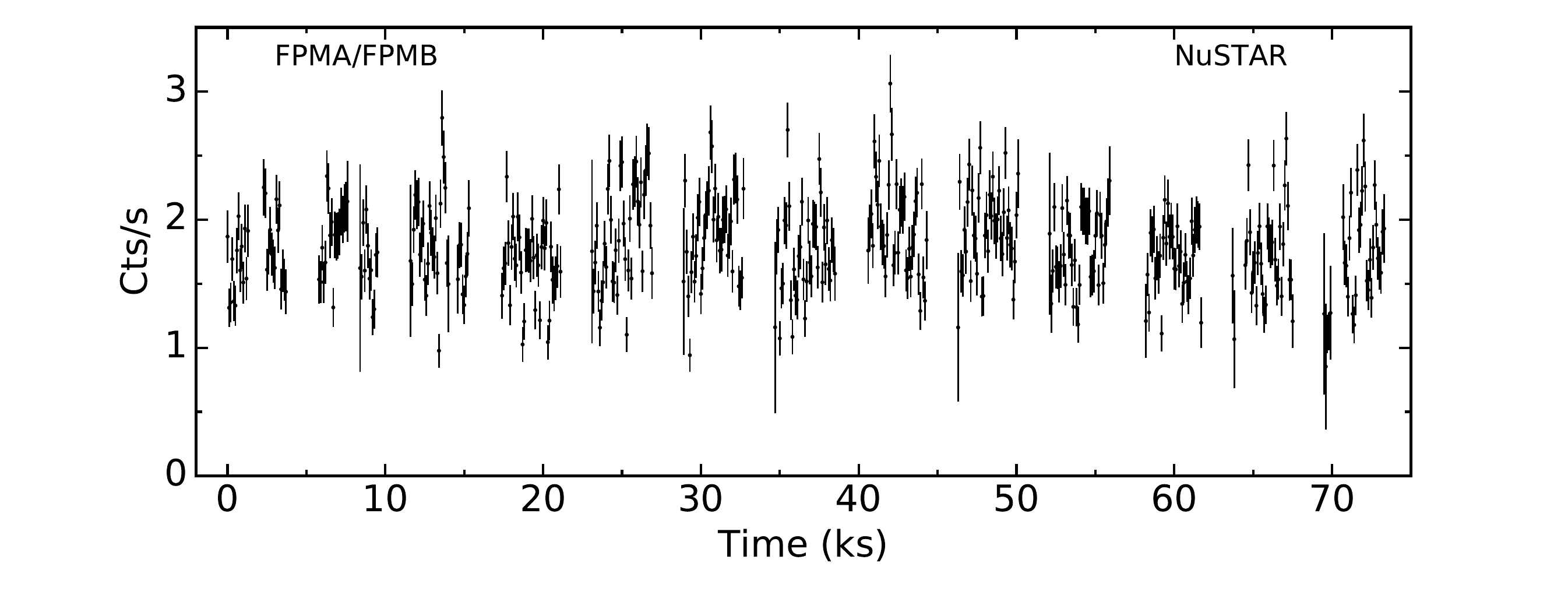}\label{fig:nulc}}
\subfigure[]{\includegraphics[width=\columnwidth]{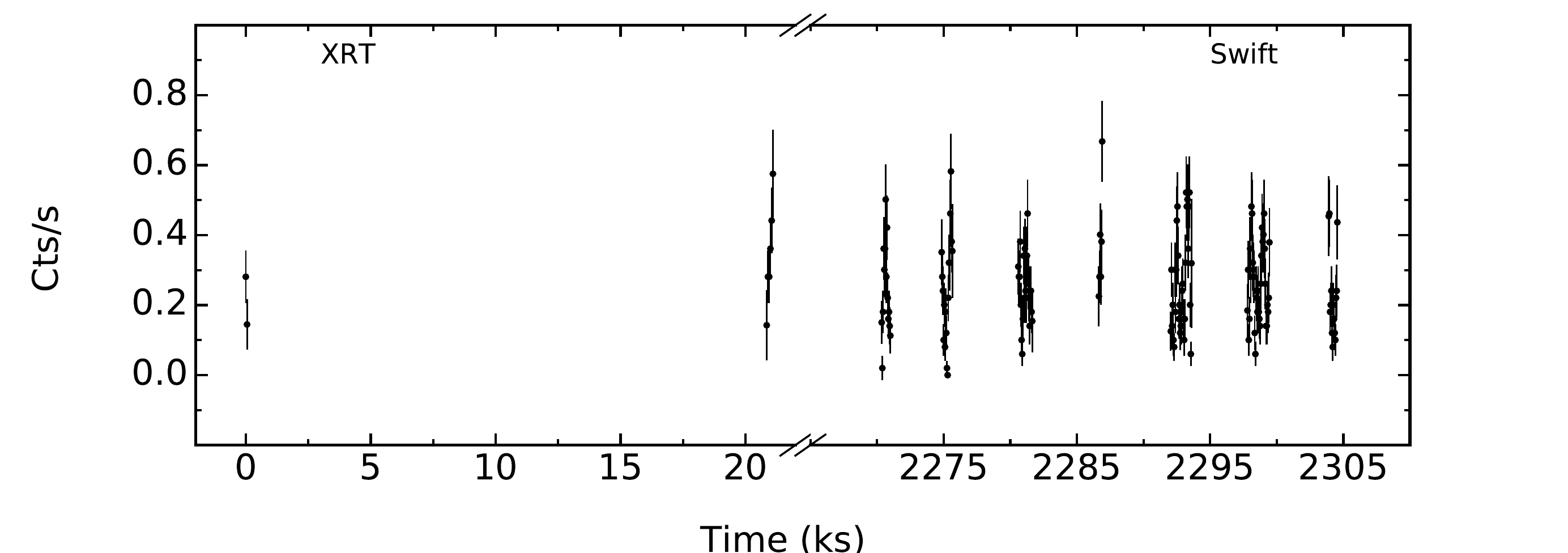}\label{fig:swlc}}
  \caption{X-ray light curves of IGR1509 as obtained from the (a) 
  \textit{Suzaku}, (b) \textit{NuSTAR}, and (c) \textit{Swift} observations in the 0.3-10.0 keV (XIS-FI/BI), 12-50 keV (HXD-PIN), 3-78 keV (FPMA/FPMB), and 0.3-10.0 keV (XRT) energy bands.}
\label{fig:xraylc}
\end{figure}

\begin{figure*}
\centering
\subfigure[]{\includegraphics[width=\columnwidth]{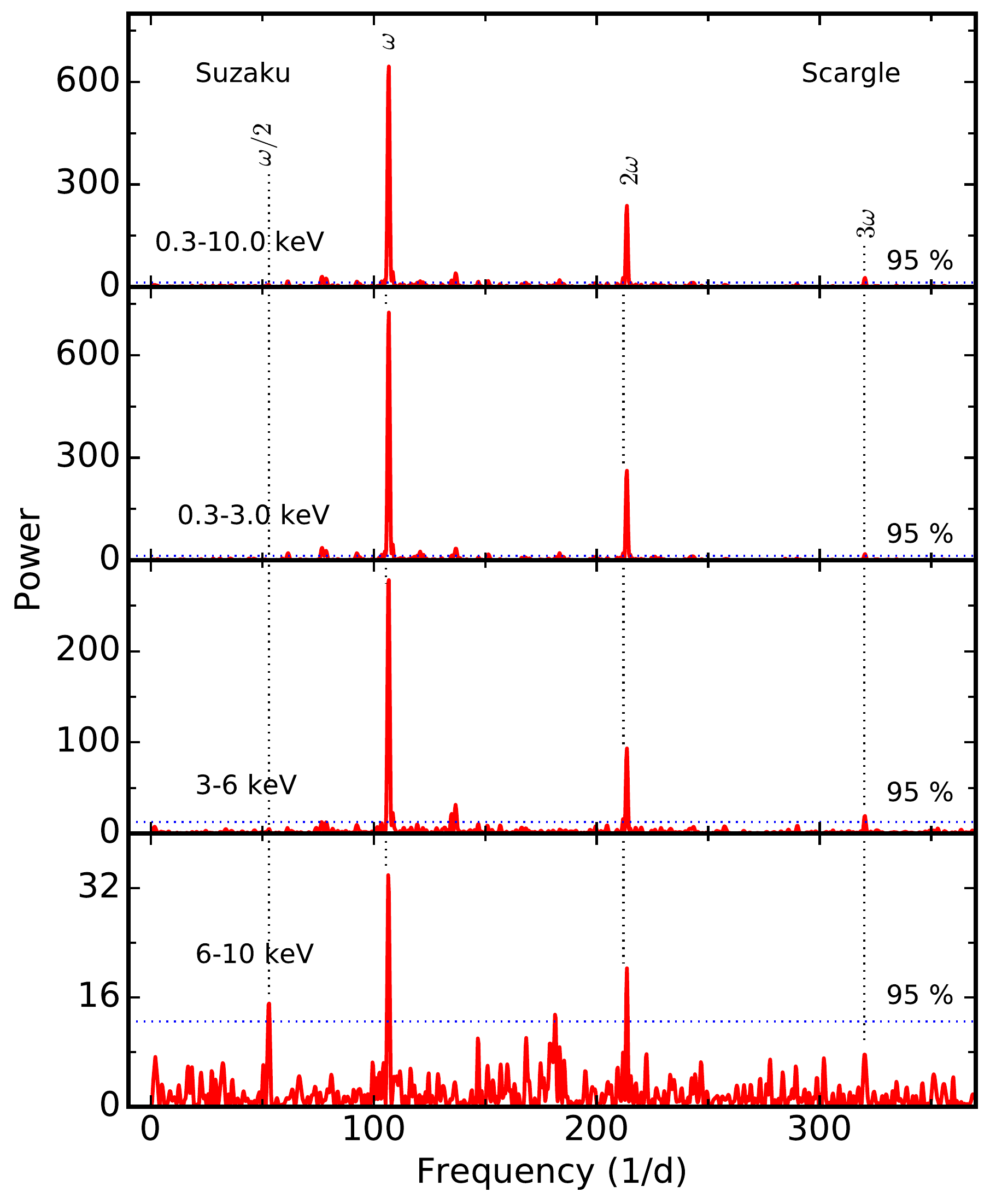}\label{fig:szscps}}
\subfigure[]{\includegraphics[width=\columnwidth]{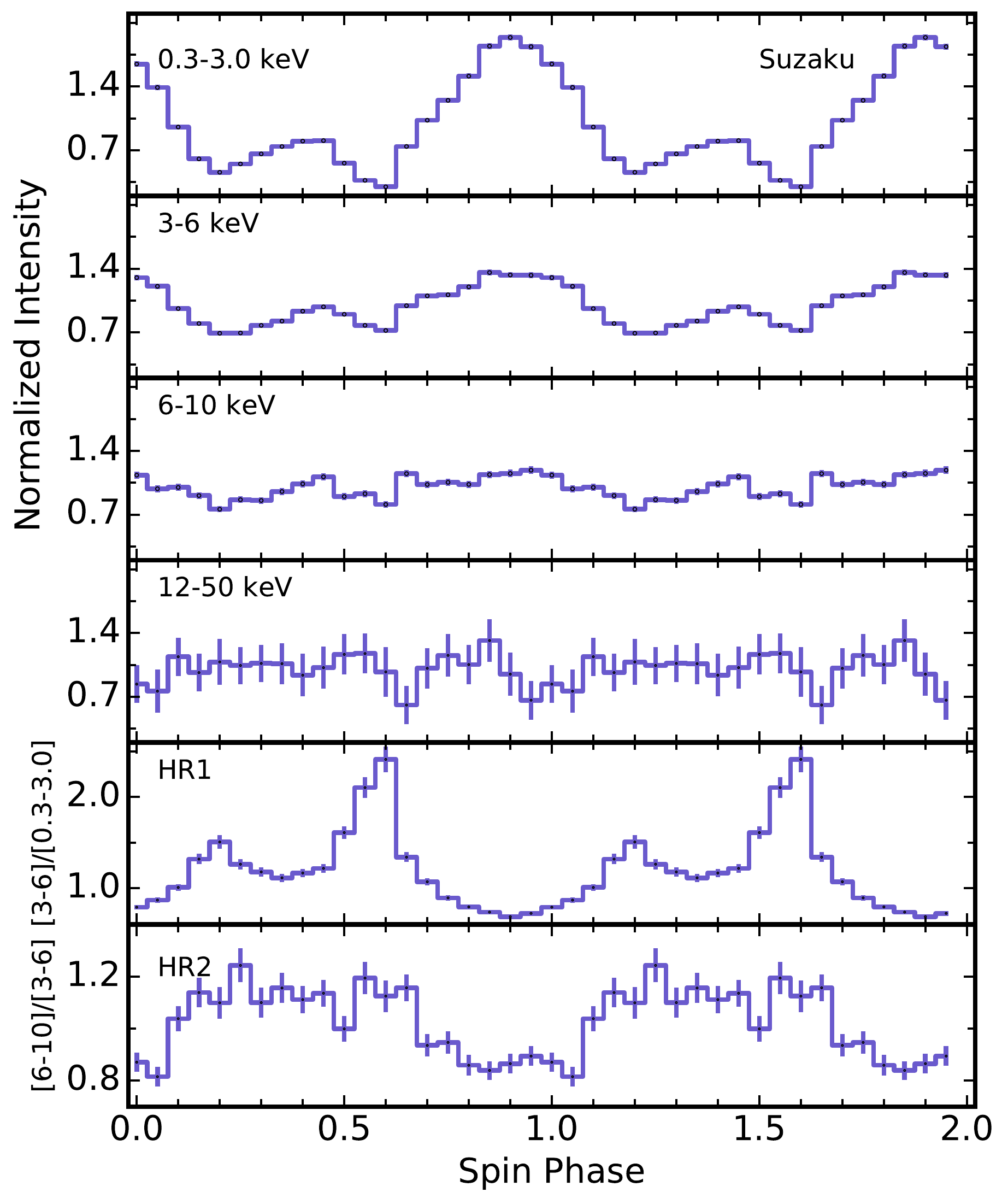}\label{fig:szflc}}
\caption{(a) Top to bottom panel shows the LS X-ray power spectra obtained from the XIS-FI observations of the \textit{Suzaku} in the 0.3-10.0 keV, 0.3-3.0 keV, 3-6 keV, and 6-10 keV energy bands, respectively. The horizontal dotted lines represent the 95\% confidence level. (b) Energy-dependent X-ray spin pulse profiles extracted from the XIS-FI observations in the 0.3-3.0 keV, 3-6 keV, and 6-10 keV energy bands and from the HXD-PIN observations in the 12-50 keV energy band. Bottom two panels represent the hardness ratio curves HR1 (=(3-6)/(0.3-3.0)) and HR2 (=(6-10)/(3-6)) obtained from the XIS-FI observations of IGR1509.}
\end{figure*}

\subsubsection{\textit{NuSTAR}}
IGR1509 was observed with the hard X-ray focusing observatory 
\textit{NuSTAR} \citep{Harrison13} on 19 July 2018 at 23:24:37 (UT) for the exposure time of 41.3 ks with an offset of 2.139 arcmin. \textit{NuSTAR} consists of two co-aligned telescopes and the two focal plane modules, FPMA and FPMB, and observes in the 3-79 keV energy range. The standard 
\textit{NuSTAR} Data Analysis Software (\textsc{NSuTARDAS v1.4.1}) was used for the data reduction. The unfiltered events were first reprocessed by using \textsc{nupipeline} in the presence of the updated version of Calibration data files (CALDB 20191219) and then the science quality events were obtained after reprocessing. The \textit{NuSTAR} source light curves and spectra were extracted by selecting a circular region of $70\arcsec$ around the source position. We used different extraction regions for FPMA and FPMB modules of the \textit{NuSTAR}, based on respective images of each module, to consider the relative astrometric offset between them. To avoid contamination from the source photons, the background light curves and spectra were accumulated by considering the same size circular region centered around four arcmin away from the source and located on the same detector chip as the source. The barycentric corrected light curves, spectra, effective area files, and response matrices were obtained via \textsc{nuproducts} package. All spectra were binned with a minimum of 20 counts per energy bin.

\begin{figure*}
\centering
\subfigure[]{\includegraphics[width=\columnwidth]{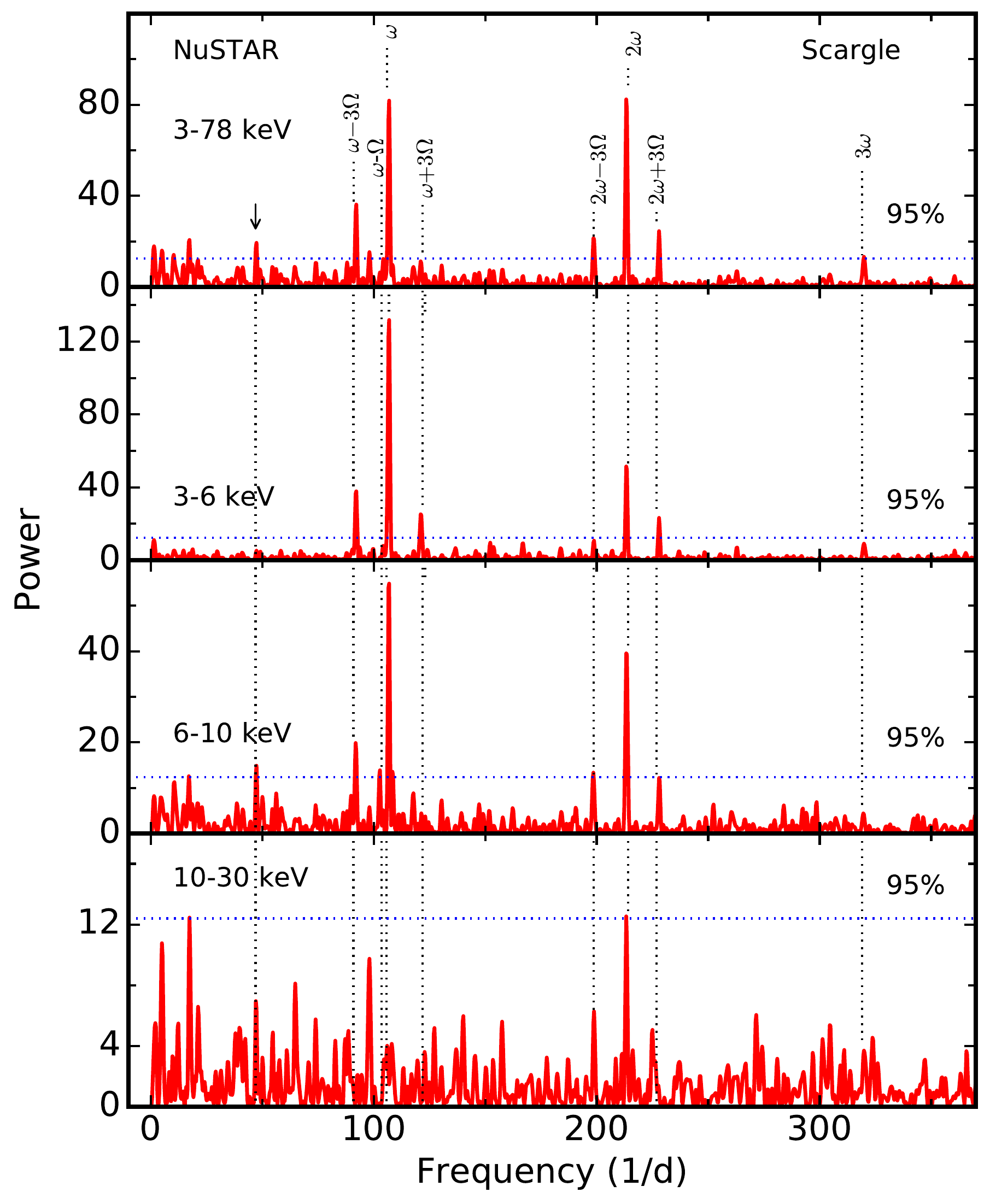}\label{fig:nuscps}}
\subfigure[]{\includegraphics[width=\columnwidth]{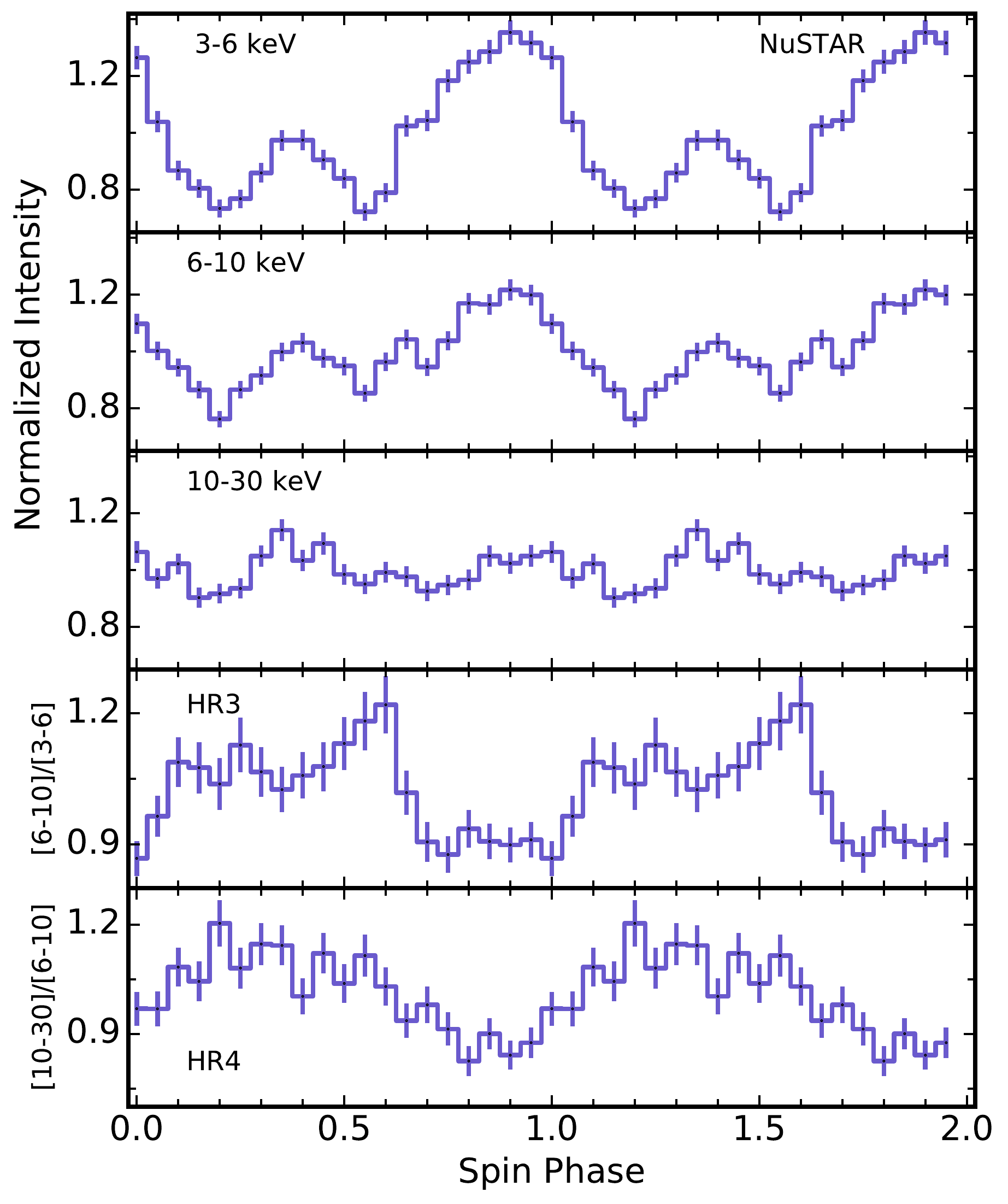}\label{fig:nuflc}}
\caption{(a) Top to bottom panel shows the LS X-ray power spectra as obtained from the \textit{NuSTAR} observations in the 3-78 keV, 3-6 keV, 6-10 keV, and 10-30 keV energy bands, respectively. The horizontal dotted lines represent the 95\% confidence level. (b) Energy-dependent X-ray spin pulse profiles in the 3-6 keV, 6-10 keV, and 10-30 keV energy bands and hardness ratio curves HR3 (=(6-10)/(3-6)) and HR4 (= (10-30)/(6-10)) obtained from the \textit{NuSTAR} observations of IGR1509.}
\end{figure*}

\subsubsection{\textit {Swift}}
IGR1509 has been observed with the \textit{Swift} satellite on 3 occasions $-$ 23 June 2018 at 17:05:42 (UT), 19 July 2018 at 23:45:03 (UT), and 20 July 2018 at 00:59:57 (UT) with offsets of 1.907 arcmin, 3.279 arcmin, and 3.084 arcmin, respectively. For these three epochs of observations, the exposure times were 0.3 ks, 0.7 ks, and 6 ks, respectively. We excluded observations of the epoch 23 June 2018 due to their short exposure and used observations of two successive epochs 19 and 20 July 2018 for further timing and spectral analyses. The \textit{Swift} consists of three instruments: the wide-field Burst Alert Telescope \citep[BAT;][]{Barthelmy05}, which covers 15-350 keV energy range; the narrow-field instruments: the X-ray Telescope \citep[XRT;][]{Burrows05} observes in 0.3-10.0 keV energy range, and the UV/Optical Telescope \citep[UVOT;][]{Roming05} with filters covering 1700-6500 \AA. The task \textsc{xrtpipeline} (version 0.13.4) along with the latest calibration files were used to produce the cleaned and calibrated event files. The barycentric corrected source light curves and spectra were extracted by selecting a circular region of 60$\arcsec$ radius. The background was chosen from a nearby source-free region with a similar size to that of the source. An ancillary response file 
  was also calculated for correcting the loss of the counts due to hot columns and bad pixels using exposure maps with the task \textsc{xrtmkarf} and used the response matrix file, $swxpc0to12s6$\_2$0130101v014.rmf$ provided by the \textit{Swift} team. All spectra from the \textit {Swift}/XRT were grouped with a minimum of 20 counts per bin.

\begin{figure*}
\centering
\subfigure[]{\includegraphics[width=\columnwidth]{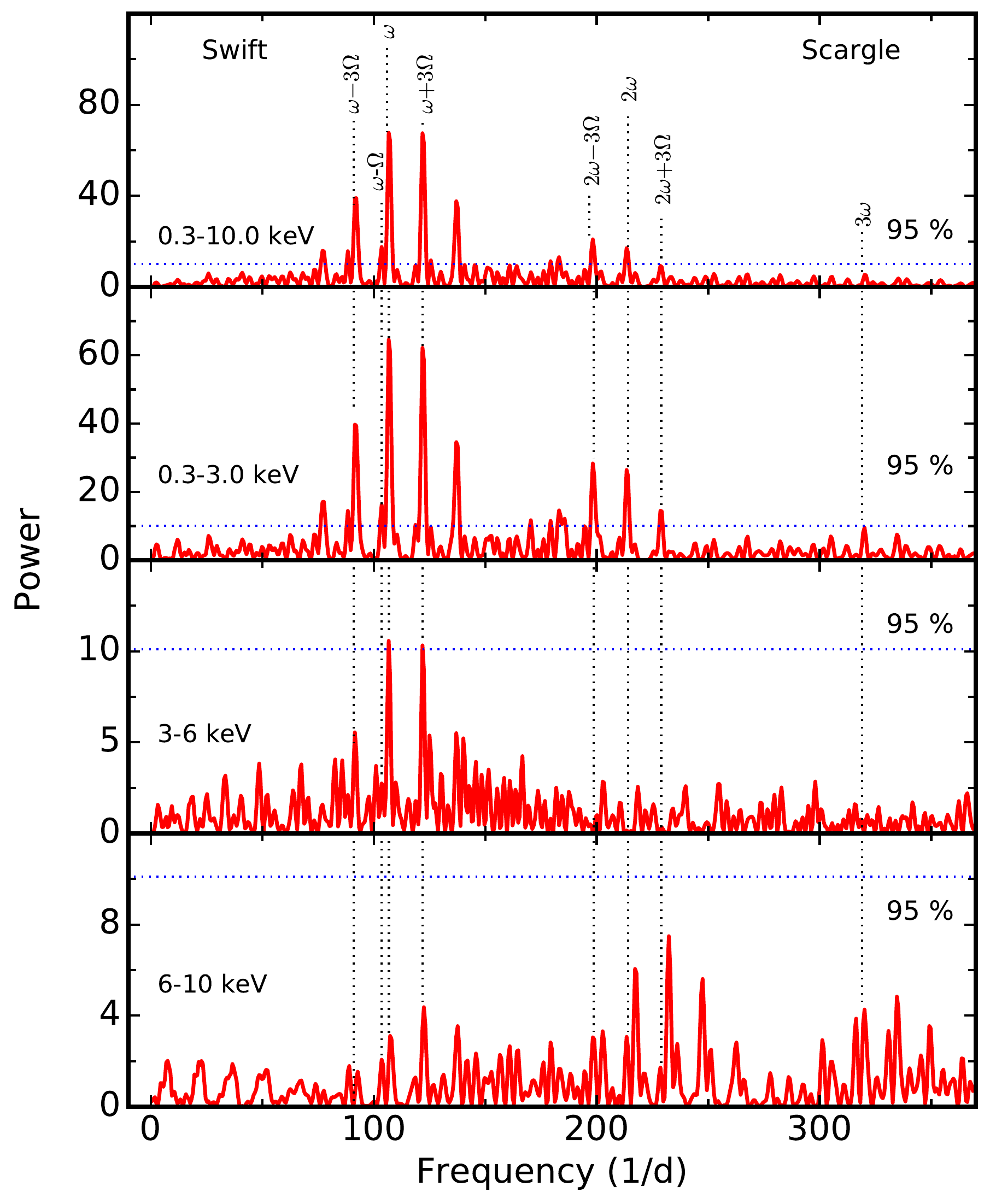}\label{fig:swscps}}
\subfigure[]{\includegraphics[width=\columnwidth]{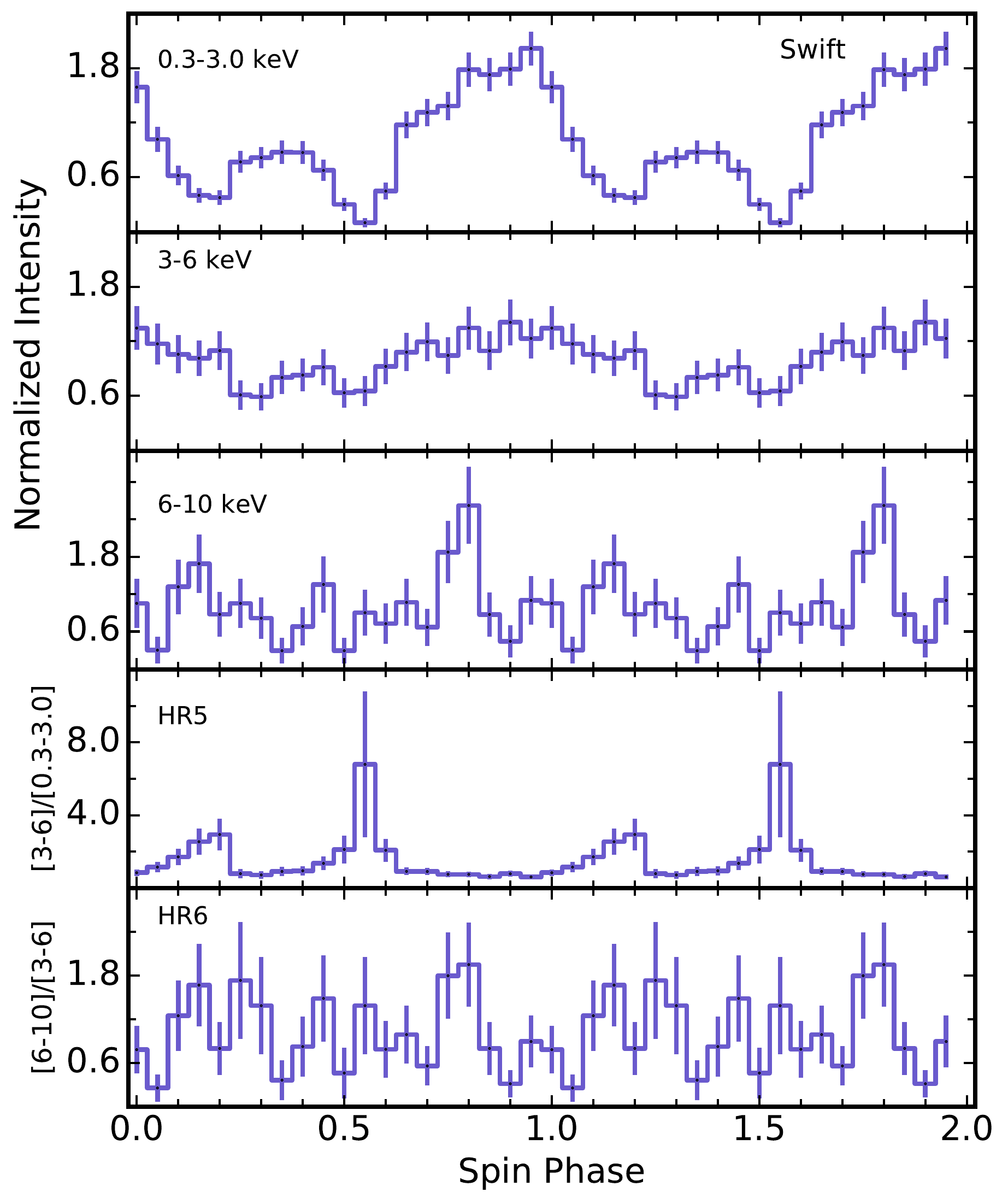}\label{fig:swflc}}
\caption{(a) Top to bottom panel shows the LS X-ray power spectra as obtained from the \textit{Swift} observations in the 0.3-10.0 keV, 0.3-3.0 keV, 3-6 keV, and 6-10 keV energy bands, respectively. The horizontal dotted lines represent the 95\% confidence level. (b) Energy-dependent X-ray spin pulse profiles in the 0.3-3.0 keV, 3-6 keV, and 6-10 keV energy bands and hardness ratio curves HR5 (=(3-6)/(0.3-3.0)) and HR6 (=(6-10)/(3-6)) obtained from the 
\textit{Swift} observations of IGR1509.}
\end{figure*}

\section{Analysis And Results} \label{sec:analysis}
\subsection{\textit {TESS}}
\subsubsection{Light curves and power spectra} \label{sec:tesslcpsflc}
Fig. \ref{fig:longtermlc} shows the AAVSO, ASAS-SN, and \textit{TESS} light curves of IGR1509 where the variable nature of the source is clearly evident. To probe IGR1509 in more detail in the optical band, we have closely inspected its temporal properties using the long-term continuous \textit{TESS} observations. For better representation, the \textit{TESS} light curves of IGR1509 are binned with a bin size of 0.025 for sectors 12, 38, and 39 (see Fig. \ref{fig:tessfulllc}).  The zoomed-in version of the two continuous days observations of sectors 12, 38, and 39 are shown in Fig. \ref{fig:tesspartlc}. The periodic behavior of the light curves was searched by performing the Lomb-Scargle \citep[LS;][]{Lomb76, Scargle82} periodogram algorithm for all three sectors separately and are shown in Fig. \ref{fig:tessscps}. We have detected six dominant peaks corresponding to the frequencies $\Omega$, $\omega$-$\Omega$, $\omega$, $2\omega$, 3($\omega$$-$$2\Omega$), and $3\omega$ in the LS power spectrum of each sector. The significance of these detected peaks was obtained by calculating the false alarm probability \citep{Horne86}. All detected peaks are found to be significant above 95\% confidence level. Periods corresponding to these six dominant peaks were also derived from the combined \textit{TESS} data of all three sectors, which are well consistent with the periods derived from each sector and are given in Table \ref{tab:ps}. 

We have also inspected the periodic variability by folding the \textit{TESS} light curves using the time of the first AAVSO-2007 observation, BJD=2454223.434948911 as the reference epoch and spin period of 809.49584 s, which is derived more precisely from the combined \textit {TESS} observations of all three sectors (see Table \ref{tab:ps}). The 
\textit{TESS} spin-phase-folded light curves with phase bin of 0.02 are shown in Fig. \ref{fig:tessspinflc}, which reveals a broad hump during a rotation of the WD in IGR1509.

\subsubsection{One-day time-resolved \textit{TESS} power spectra and phased light curves} \label{sec:1dayps}
Using the unique capability of the long baseline of the \textit{TESS}, we have inspected the evolution of the power spectra in consecutive one-day time segments. Each sector data was divided into consecutive one-day segments. In this way, we have a total of 27, 26, and 27 data segments for sectors 12, 38, and 39, respectively. Each one-day data was further used for LS periodogram analysis. Figs. \ref{fig:tess1daypssec12}, \ref{fig:tess1daypssec38}, and \ref{fig:tess1daypssec39} show trailed power spectra with a 1-day increment for sectors 12, 38, and 39, respectively. A significant spin peak is detected in each day's observation of all three sectors. However, the $\Omega$ and $2\omega$ frequencies are sporadic throughout the time-resolved power spectra, which are sometimes significant and sometimes lie below the confidence level. But no significant $3\omega$ and $3(\omega-2\Omega)$ frequencies were detected in one day power spectra of IGR1509. In contrast to combined power spectra, a significant $\omega-\Omega$ signal is not detected during one-day observations of IGR1509 for all three sectors.  However, as the bin size increases up to the minimum of seven consecutive days or beyond that, the beat frequency was found to be present with a 90\% confidence level for all three sectors 12, 38, and 39.
 
Further, to see the evolution of the short-term variations, we have also explored the day-wise periodic variation during the rotation of the WD. Using a similar approach as described in Section \ref{sec:tesslcpsflc}, we have folded each day's light curve with the binning of 20 points in a phase. Left to right panels in Fig. \ref{fig:tess1dayflc} represent the color composite plots for the one-day spin-phased pulse profile for all three sectors. Strong single broad-peak spin modulations are observed throughout one-day \textit{TESS} observations in all sectors, which is consistent with the time-resolved power spectrum. 

\subsection{\textit{Suzaku}} \label{sec:szanalysis}
\subsubsection{Light curves and power spectra} 
\label{sec:szlcps}
Background-subtracted X-ray light curves of IGR1509 were 
 obtained from the XIS-FI, XIS-BI, and HXD-PIN instruments with the temporal binning of 16 s, 16 s, and 128 s, respectively which are shown in Fig. \ref{fig:szlc}. We have performed a period analysis by applying the LS method to the XIS-FI  data. Fig. \ref{fig:szscps} shows the LS power spectra of the XIS-FI data in the 0.3-10.0 keV, 0.3-3.0 keV, 3-6 keV, and 6-10 keV energy bands. Similar to the optical \textit {TESS} power spectra, the prominent spin frequency is detected in its all-energy band X-ray power spectra. Along with the $\omega$, two other peaks corresponding to the frequencies $2\omega$ and $3\omega$ are also found to be present. All detected peaks are significant and lie above 95\% confidence level excluding $3\omega$ peak in the hard 6-10 keV energy band. The peak power of these significant peaks is found to decrease towards the harder energy bands. 
   Apart from these peaks, a significant frequency corresponding to the period of $P_{\omega/2}$=1627.2$\pm$8.9 s is also detected in the 6-10 keV energy band power spectrum. The derived significant periods in the 0.3-10.0 keV energy band are given in Table \ref{tab:ps}. 

\subsubsection{Energy-dependent periodic variations} \label{sec:szenflc}
The periodic variability was also inspected in three energy bands: 0.3-3.0 keV, 3-6 keV, and 6-10 keV using XIS-FI observations. We have folded the XIS-FI light curves using the same epoch as mentioned above in Section \ref{sec:tesslcpsflc}  and by considering accumulated timing uncertainty over a spin period of 809.49584 s. Energy-dependent X-ray spin pulse profiles obtained from the XIS-FI observations are shown in Fig. \ref{fig:szflc}. We have also extracted the spin-phase-folded light curves for the HXD-PIN observations in the 12-50 keV energy band. No significant spin modulations are detected in the 12-50 keV energy band (see Fig. \ref{fig:szflc}). Spin-phase-folded light curves are found to be energy-dependent and appear to be decreasing in amplitude as moving toward hard energies. This is also evident from the power spectrum where the peak power of the spin frequency is found to be decreasing towards harder energy bands. In each phased light curve, double-peaked spin modulations are clearly seen which are found to be more prominent at lower energies and almost constant at higher energies. We have also extracted the spin-phase-folded hardness ratio curves, HR1 and HR2, where HR1 is the ratio of the count rates in 3-6 keV to count rates in 0.3-3.0 keV energy bands, and HR2 is the ratio of the count rates in 6-10 keV to count rates in 3-6 keV energy bands and are shown in the bottom two panels of Fig. \ref{fig:szflc}. The HR1 and HR2 curve displays a strong modulation and is 180$^{\circ}$ out of phase with respect to the intensity modulation, \textit {i.e.,} the maximum in the hardness ratio curve is observed at the lowest intensity. We have also estimated the degree of spin pulsations with [($I_{max}$-$I_{min}$)/($I_{max}$+$I_{min}$)]$\times$ 100\%, where $I_{max}$ and $I_{min}$ are maximum and minimum intensities in a pulse profile, respectively. The derived values of the pulsed fraction are given in Table \ref{tab:fracmod}, which clearly shows a decreasing trend as the energy increases. It decreased from 73$\pm$2\% in the 0.3-3.0 keV energy band to 22$\pm$3\% in the 6-10 keV energy band. However, no significant periodic variation was detected in the 12-50 keV energy band.
\begin{table*}
\centering
\small
\caption{Energy dependent X-ray spin modulation obtained from the \textit {Suzaku}, \textit{NuSTAR}, and \textit{Swift} observations.\label{tab:fracmod}}
\setlength{\tabcolsep}{0.05in}
\begin{tabular}{lcccccccccc}
\hline
Telescope    & Epoch & \multicolumn{5}{c}{Spin Modulation (\%)}           \\
\cline{3-7}\\
             &       &  0.3-3.0 keV & 3-6 keV  & 6-10 keV & 10-30 keV & 12-50 keV \\
\hline\\
\textit{Suzaku}      &  2011  &73$\pm$2       &33$\pm$2 &22$\pm$3 &\nodata   & \nodata\\ 
\textit{NuSTAR}      &  2018  &\nodata        &31$\pm$3 &23$\pm$2 & 12$\pm$3 &  \nodata  \\   
\textit{Swift}      &  2018  &91$\pm$13      &41$\pm$16&\nodata  & \nodata  &  \nodata   \\ 
\hline                                                                     \end{tabular}
\end{table*}

\subsection{\textit{NuSTAR}} \label{sec:nuanalysis}
\subsubsection{Light curves and power spectra} \label{sec:nulcps}
Background-subtracted X-ray light curves were obtained from the FPMA and FPMB instruments of the \textit{NuSTAR} with the temporal binning of 10 s. Both FPMA and FPMB light curves were combined using the task 
\textsc{lcmath} and the combined light curve in the 3-78 keV energy band is shown in Fig. \ref{fig:nulc}. Using the LS method, we performed the periodogram analysis in the 3-78 keV, 3-6 keV, 6-10 keV, and 10-30 keV energy bands. The \textit{NuSTAR} LS power spectra are shown in Fig. \ref{fig:nuscps}. We have detected three significant periods corresponding to the frequencies $\omega$, $2$$\omega$, and $3$$\omega$ in the 3-78 keV energy band, which are well consistent with the periods derived from the \textit{TESS} and \textit{Suzaku} observations. In contrast to the \textit{Suzaku}, the marginal detection of the X-ray beat frequency also appears to be present in the \textit{NuSTAR} power spectra which is well consistent with the optical beat frequency derived from the \textit{TESS} observations. The periods corresponding to the frequencies $\omega$, $\omega$$-$$\Omega$, $2$$\omega$, and $3$$\omega$ in the 3-78 keV energy band are given in Table \ref{tab:ps}. A significant peak marked with an arrow near 47 day$^{-1}$ frequency region is also observed in the 3-78 keV and 6-10 keV energy band which is corresponding to the period of 1822.1$\pm$11.3 s. This peak has no possible relation with other 
 present frequencies found in the power spectrum. Four other periods which are a combination of the spin and orbital frequencies as $P_\omega$$_-$$_3$$_\Omega$=937.2$\pm$2.9 s, $P_\omega$$_+$$_3$$_\Omega$=713.7$\pm$1.7 s, $P_{2\omega}$$_-$$_3$$_\Omega$=435.2$\pm$0.6 s, and $P_{2\omega}$$_+$$_3$$_\Omega$=379.0$\pm$0.5 s are also seen in the \textit{NuSTAR} power spectra. 

\subsubsection{Energy-dependent periodic variations} \label{sec:nuenflc} 
Using the approach as described in Section \ref{sec:szenflc}, we have generated spin-phased light curves in the 3-6 keV, 6-10 keV, and 10-30 keV energy bands using the \textit{NuSTAR} observations. All folded light curves were extracted with the temporal binning of 20 s and are shown in Fig. \ref{fig:nuflc}. The \textit{NuSTAR} spin-phase-folded light curves also exhibit a double-peaked pulse profile which are more pronounced in the softest energy bands. We have also extracted, the spin-phase-folded hardness ratio curves, HR3 and HR4, where HR3 is defined as the ratio of the count rates in 6-10 keV to count rates in 3-6 keV energy bands, and HR4 is the ratio of the count rates in 10-30 keV to count rates in 6-10 keV energy bands and are shown in the bottom two panels of Fig. \ref{fig:nuflc}. HR3 and HR4 curves exhibit anti-correlated modulations with respect to the intensity modulation. The fractional amplitude of modulation was also derived from the \textit{NuSTAR} folded light curves which are given in Table \ref{tab:fracmod}. The amplitudes of modulation are decreasing towards higher energies, which decreased from 31$\pm$3\% in the 3-6 keV energy band to 12$\pm$3\%  in the 10-30 keV energy band.

\subsection{\textit {Swift}} \label{sec:swanalysis}
\subsubsection{Light curves and power spectra} \label{sec:swlcps}
Fig. \ref{fig:swlc} shows the X-ray light curves of IGR1509 in the 0.3-10.0 keV energy band as extracted from two successive epochs 19 and 20 July 2018 of the \textit{Swift}-XRT observations. The \textit{Swift} LS power spectra computed from the light curves with a time bin of 10 s in the 0.3-10.0 keV, 0.3-3.0 keV, 3-6 keV, and 6-10 keV energy bands and are shown in Fig. \ref{fig:swscps}. Four significant periods at frequencies $\omega$, $\omega$$-$$\Omega$, $2\omega$, and $3\omega$ are detected in the \textit{Swift} LS power spectra, which are well consistent with the periods derived from the \textit{TESS}, \textit{Suzaku}, and \textit{NuSTAR} observations and are given in Table \ref{tab:ps}. Similar to the \textit{NuSTAR}, the other side-band frequencies corresponding to the periods $P_\omega$$_-$$_3$$_\Omega$=937.3$\pm$6.4 s, $P_\omega$$_+$$_3$$_\Omega$=709.0$\pm$3.6 s, $P_{2\omega}$$_-$$_3$$_\Omega$=435.8$\pm$1.4 s, and $P_{2\omega}$$_+$$_3$$_\Omega$=378.0$\pm$1.0 are also present in the \textit{Swift} LS power spectra of IGR1509.

\subsubsection{Energy-dependent periodic variations} \label{sec:nuenflc}
Spin-phased \textit{Swift} X-ray light curves in the 0.3-3.0 keV, 3-6 keV, and 6-10 keV energy bands are shown in Fig. \ref{fig:swflc}. The spin pulsations obtained from the \textit{Swift} data are also found to be energy-dependent, where strong modulations are observed in the softest energy band. The spin-phase-folded hardness ratio curves, HR5 and HR6, defined similar to HR1 and HR2 were also derived and are shown in the bottom two panels of Fig. \ref{fig:swflc}. The variation of the HR5 curve is anti-correlated with intensity modulation, while, no significant variations are seen in the HR6 curve and remain constant over the entire spin cycle. We have also derived the value of the degree of pulsations from the \textit{Swift} light curves. In the \textit{Swift} spin-phased light curves, the fractional amplitude modulations decreased from 91$\pm$13\% (0.3-3.0 keV) to 41$\pm$16\% (3-6 keV) energy bands. However, no significant periodic variation was detected in the 6-10 keV energy band.

\begin{figure*}
\centering
\includegraphics[width=150mm, height=115mm]{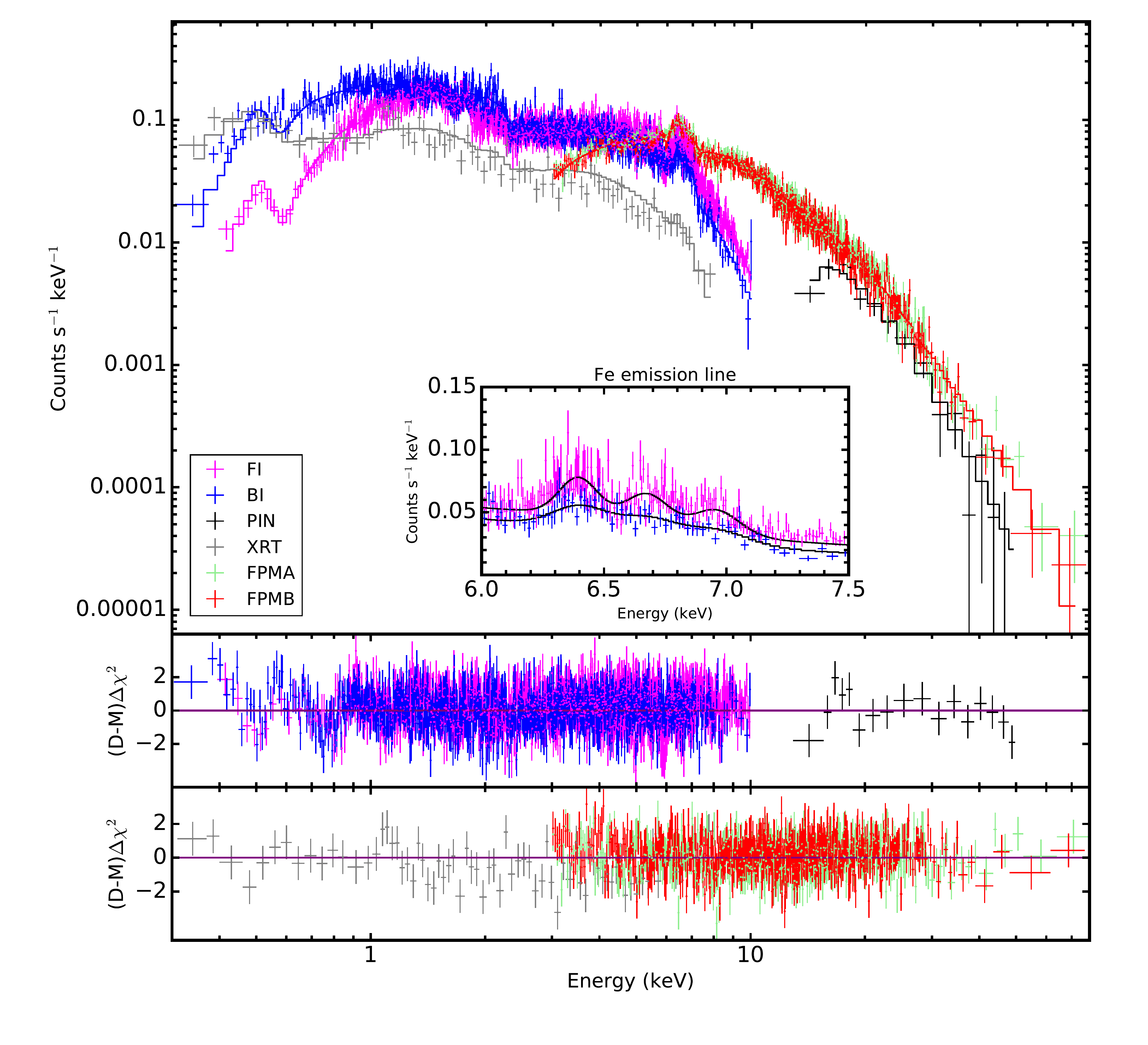}
\caption{Best-fitted X-ray spectra obtained from the \textit{Suzaku}-XIS/(FI-magenta; BI-blue) and PIN (black) in the 0.3-10.0 keV and 12-50 keV energy bands, respectively. The zoomed \textit{Suzaku}-FI and BI spectra around the Fe-line region are shown in the inset. The best-fitted combined 
\textit{Swift} (XRT-grey) and \textit{NuSTAR} (FPMA-green; FPMB-red) spectra in the 0.3-10.0 keV and 3-78 keV energy bands, respectively are also shown in the same plot. The bottom panel shows the spectral residuals obtained after fitting the broad-band spectra with respect to best-fitted model C (see in text for more detail).}
\label{fig:szswnuspec}
\end{figure*}

\subsection{X-ray spectral analysis} \label{sec:xrayspec}
\subsubsection{\textit{Suzaku} spectral fits} \label{sec:szspec}
The background-subtracted X-ray spectra obtained for the epoch 2011 of the \textit{Suzaku}-FI observations in the 0.3-10.0 keV energy band and the \textit{Suzaku}-HXD/PIN observations in the 12-50 keV energy bands are shown in Fig. \ref{fig:szswnuspec}. The strong broadband continuum along with the strong complex iron emission line features are clearly seen. The inset in Fig. \ref{fig:szswnuspec} shows an enlarged view of the 6.0$-$7.5 keV energy range, which contains the line features at 6.4 keV (Fe K$\alpha$), 6.7 keV (Fe XXV ), and 6.95 keV (Fe XXVI ). The X-ray spectral analysis was performed using XSPEC version-12.12.0 \citep{Arnaud96, Dorman01}. The spectral fitting in the present study was accomplished by models that were approximately the same as those utilised by \cite{Bernardini12}  for the \textit {XMM-Newton} data. We only replaced the interstellar absorption model \texttt {wabs} with the updated \texttt{phabs} model and the pcfabs model with the (more appropriate) \texttt{pwab}, which is a power-law distribution of a covering fraction as a function of the maximum equivalent hydrogen column $N_{H,max}$ and the power-law index for the covering fraction $\beta$ \citep{Done98}. In addition, we replace the  \citep[{\texttt mekal};][]{Mewe85} with \texttt{mkcflow} \citep{Mushotzky88}, a multi-temperature cooling flow model that is more appropriate for accreting WDs \citep[see][]{Mukai17} along with a Gaussian component with fixed-line energy and line width at 6.4 keV and 0.01 keV, respectively. To know the cross-calibration uncertainties of the distinct instruments, a constant model component was also included. We have fixed the PIN cross-normalization constant at 1.16 for the XIS-nominal pointing position (see \textit{Suzaku} Memo 2008$-$06\footnote{\textcolor{magenta}{\url{ftp://legacy.gsfc.nasa.gov/suzaku/doc/xrt/suzakumemo-2008-06.pdf}}}). The abundance tables and the photoelectric absorption cross section `bcmc' were taken from \cite{Asplund09} and \cite{Balucinska-Church92}, respectively. The redshift required in the mkcflow model cannot be zero. It was thus fixed to a value of 2.55$\times$10$^{-7}$ for a Gaia distance of 1091$_{-23}^{+25}$ pc \citep{Bailer-Jones21}. The low temperature of the mkcflow was fixed to the minimum value allowed by the model as 80.8 eV. We assumed a switch parameter at the value of 2 which determines whether the spectrum is computed by using the AtomDB data. We have fixed an equivalent hydrogen column from the phabs model to the total Galactic column in the direction of IGR1509, that is, at 1.89$\times$10$^{21}$ cm$^{-2}$ given by the HI4PI survey \citep{HI4PICollaboration16}. Moreover, to account for the excess in the soft X-rays, we have used an additional blackbody component as A=constant$\times$phabs$\times$pwab(bb+mkcflow+gauss). With model A, we have found that the spectrum fitted well with the $\chi^2_\nu$ of 1.06. Along with model A, we have obtained the large equivalent width (EW $>$ 100 eV) of the Fe K$\alpha$ emission line and the column density of cold matter is found to be somewhat higher than 2$\times$10$^{23}$ cm$^{-2}$ \citep[see][]{Inoue85}. This suggests that the bulk of the Fe K$\alpha$ emission line can be accounted by the absorbers, while there may also be a contribution of a Compton reflected continuum from the WD, which is expected to go along with the fluorescent iron line \citep[see][]{Matt91, Done92, Ezuka99}. A strong Fe K$\alpha$ emission line with a large EW of 170 eV was also detected by \cite{Bernardini12} using the \textit{XMM-Newton} data. But their data did not allow them to restrict the reflection component. Therefore, to take into account the reflected continuum of the plasma emission due to the Fe K$\alpha$ at 6.4 keV, we used a convolution model \texttt{reflect} with the fixed inclination angle of the reflecting surface at the default value of cos$(i)$= 0.45 as model B=constant$\times$phabs$\times$pwab(bb+reflect$\times$mkcflow+gauss). Using model B, we have found a better fit near the iron line complex region, but the value of the reflection scaling factor ($R_{refl}$) was derived to be more than unity. The reflection component describes the fraction of downward radiation that is reflected, so, the observed value of more than unity is not physical. Thus, we fixed this parameter to unity allowing cos$(i)$ to vary as model C, and we obtained fits with a similar quality as the best-fit value cos$(i)$ $>$ 0.30. Unabsorbed X-ray flux in the 0.3-50.0 keV energy band was also calculated using the `\texttt{cflux}' model. The best-fit parameters as obtained for each model are given in Table \ref{tab:xrayspec}, where the error bars are quoted with a 90\% confidence limit for a single variable parameter.

\begin{table*}
\small
\begin{center}
\caption{Spectral parameters obtained from the best-fit models to the average \textit{Suzaku} (epoch 2011), and contemporaneous \textit{Swift} and \textit{NuSTAR} (epoch 2018) observations.\label{tab:xrayspec}}
\setlength{\tabcolsep}{0.015in}
\begin{tabular}{cccccccccccccccccccccc}
\hline
 Epoch            &             &     & \multicolumn{2}{c}{pwab}       && \multicolumn{2}{c}{blackbody} && \multicolumn{2}{c}{reflect}    && \multicolumn{3}{c}{mkcflow}                  & \multicolumn{3}{c}{gaussian}        &X-ray Flux                & $\chi^2_\nu$ (dof)  \\
\cline{4-5} \cline{7-8} \cline{10-11} \cline{13-15} \cline{17-18}
                  & Model($\downarrow$) &     & $N_{H, max}$&  $\beta$           && $kT_{bb}$&  $n_{bb}$          && $R_{refl}$ &$Cosi$             && $kT$      & $A_{z}$   & $n_{mkcflow}$          && $n_{g6.4}$  & $EW_{g6.4}$          & $f_{X}$                  &  \\
\hline\\
  2011$^\dagger$  &A     	     &     & $>$ 5.6     & -0.75$_{-0.01}^{+0.07}$ && 97$_{-3}^{+3}$  &3.92$_{-0.62}^{+5.46}$&& ..                  &..           &&$>34$                 &0.63$_{-0.29}^{+0.13}$ &4.34$_{-0.35}^{+9.09}$   && 2.86$_{-0.42}^{+1.05}$ & 126$_{-10}^{+32}$   &7.16$_{-0.04}^{+0.04}$    & 1.06(3093)  \\
		  &B	             &     & 2.0$_{-0.6}^{+0.9}$      & -0.75$_{-0.02}^{+0.02}$ && 87$_{-3}^{+3}$  &2.14$_{-0.52}^{+0.65}$&&5.4$_{-1.6}^{+1.8}$  &0.45$^*$     &&30$_{-4}^{+5}$        &0.35$_{-0.06}^{+0.07}$ &3.97$_{-0.84}^{+1.11}$   && 1.73$_{-0.31}^{+0.38}$ & 100$_{-14}^{+29}$   &4.84$_{-0.03}^{+0.03}$    & 1.05(3093)   \\
		  &C        	     &     & $>$ 4.4      & -0.70$_{-0.02}^{+0.01}$ && 89$_{-2}^{+3}$  &4.76$_{-1.14}^{+1.72}$&&1$^*$                &$>0.30$      &&30$_{-3}^{+14}$       &0.39$_{-0.05}^{+0.17}$ &8.32$_{-3.66}^{+5.79}$  && 2.63$_{-0.32}^{+0.58}$ & 99$_{-14}^{+29}$    &5.26$_{-0.03}^{+0.03}$    & 1.04(3092)   \\
  2018$^\ddagger$ &A                 &     & 78.9$_{-22.1}^{+29.0}$   & -0.66$_{-0.02}^{+0.03}$ && 84$_{-9}^{+9}$  &29.17$_{-9.50}^{+15.54}$&&..                 &..           &&23$_{-3}^{+3}$        &0.14$_{-0.04}^{+0.04}$ &32.3$_{-8.48}^{+12.58}$  && 8.31$_{-1.56}^{+2.04}$ & 134$_{-19}^{+17}$   &24.94$_{-0.19}^{+0.18}$   & 0.94(986)  \\
		  &~B	             &     & 7.9$_{-3.0}^{+6.5}$      & -0.73$_{-0.02}^{+0.03}$ && 81$_{-9}^{+9}$  &8.96$_{-2.77}^{+4.47}$&&1.9$_{-0.7}^{+0.8}$ &0.45$^*$      &&27$_{-2}^{+3}$        &0.14$_{-0.04}^{+0.04}$ &10.2$_{-1.42}^{+2.29}$ 	 && 3.61$_{-0.69}^{+0.87}$ & 132$_{-18}^{+23}$   &9.79$_{-0.07}^{+0.07}$    & 0.95(985)  \\
		  &~C	             &     & 11.0$_{-2.6}^{+4.8}$     & -0.72$_{-0.02}^{+0.01}$ && 81$_{-9}^{+9}$  &10.43$_{-3.08}^{+5.01}$&&1$^*$                &$>0.62$     &&27$_{-3}^{+3}$        &0.14$_{-0.04}^{+0.04}$ &11.6$_{-1.74}^{+2.09}$ 	 && 4.00$_{-0.67}^{+0.76}$ & 146$_{-23}^{+16}$   &10.83$_{-0.10}^{+0.10}$   & 0.95(985)  \\
\hline
\end{tabular}
\end{center}
\textit{Notes:} {$\dagger$=\textit{XIS+PIN} and $\ddagger$ = XRT+FPMA/FPMB. $A$, $B$, and $C$ are the best-fitted models for the epochs 2011 and 2018 (see text of the Sections \ref{sec:szspec} and \ref{sec:swnuspec} for more detail of each model). $*$ represents fixed parameter, $N_{H, max}$ is the maximum equivalent hydrogen column in units of $10^{23}$ cm$^{-2}$ and $\beta$ is the power-law index for the covering fraction. $kT_{bb}$ is the blackbody temperature in units of eV, $n_{bb}$ is the normalization constant of blackbody component in units of 10$^{-4}$. $R$$_{refl}$ is reflection scaling factor and $Cosi$ is the inclination angle of the reflecting surface. $kT$ is the high temperature of mkcflow model in units of keV, $n_{mkcflow}$ is the normalization of the mkcflow model in units of 10$^{-10}$$M_\odot$ yr$^{-1}$, $n_{g6.4}$ is normalization constant of the Gaussian component in units of 10$^{-5}$, $EW_{g6.4}$ is the equivalent width of Fe K$\alpha$ in units of eV, and $f_{X}$ is the unabsorbed X-ray flux derived in the 0.3-50.0 keV and 0.3-78.0 keV energy bands in units of 10$^{-11}$ erg cm$^{-2}$ s$^{-1}$ for epochs 2011 and 2018, respectively. All the errors are within a 90\% confidence interval for a single parameter ($\Delta$ $\chi^2$=2.706)}.
\end{table*}

\begin{table*}
\tabletypesize{\tiny}
\begin{center}
  \caption{Best-fit spectral parameters derived from the spectral fitting of the \textit{Suzaku}-XIS and contemporaneous \textit{Swift}-XRT and \textit{NuSTAR}-FPMA/FPMB observations at phases of maximum (0.8-1.0) and minimum (0.5-0.7) of spin pulsation. \label{tab:szswnuspinprs}}
\setlength{\tabcolsep}{0.05in}
\begin{tabular}{ccccccccccccccccccc}
\hline
Model        & Parameters & \multicolumn{2}{c}{\textit {Suzaku}}  &&  \multicolumn{2}{c}{\textit {Swift} and \textit {NuSTAR}}  \\
\cline{3-4} \cline{6-7} \\ 
             &            &  Pulse Maximum      &   Pulse Minimum    &&    Pulse Maximum      &   Pulse Minimum \\                                                              
\hline
 pwab        & $N_{H,max}$ ($\times$10\textsuperscript{23} cm\textsuperscript{-2})                                & 2.5$_{-0.9}^{+0.3}$      & 4.6$_{-0.7}^{+1.4}$       &&  7.1$_{-3.1}^{+6.4}$  & 11.8$_{-3.3}^{+4.3}$             \\ 
                     & $\beta$                                                                                    & -0.80$_{-0.02}^{+0.02}$    & -0.48$_{-0.02}^{+0.02}$   && -0.80$_{-0.04}^{+0.05}$ &-0.64$_{-0.04}^{+0.05}$          \\ 
Mkcflow    & $n_{mkcflow}$ ($\times$10\textsuperscript{-10}) $M_\odot$ yr$^{-1}$                                  & 10.72$_{-0.02}^{+0.02}$    & 10.50$_{-0.62}^{+1.12}$   &&  16.15$_{-0.85}^{+0.10}$& 15.82$_{-0.96}^{+1.04}$            \\
Gaussian   & $n_{g6.4}$ ($\times$10\textsuperscript{-5})                                                          & 2.59$_{-0.57}^{+0.98}$     & 2.44$_{-0.80}^{+0.80}$    &&  3.69$_{-1.59}^{+1.71}$ & 5.60$_{-1.99}^{+2.05}$            \\
           & $EW_{g6.4}$  & 105$_{-29}^{+16}$ & 86$_{-19}^{+27}$ && 111$_{-46}^{+39}$ & 122$_{-24}^{+88}$ \\
X-ray flux & $f_{X0.3-10.0}$($\times$10\textsuperscript{-11}erg cm\textsuperscript{-2} s\textsuperscript{-1})     & 6.04$_{-0.06}^{+0.05}$     & 5.68$_{-0.08}^{+0.08}$    &&  $..$                   & $..$                              \\
           & $f_{X0.3-78.0}$($\times$10\textsuperscript{-11}erg cm\textsuperscript{-2} s\textsuperscript{-1})     & $..$                       & $..$                      &&  12.56$_{-0.19}^{+0.19}$& 12.40$_{-0.22}^{+0.22}$             \\    
           & $\chi_\nu^2$ (dof)                                                                                   & 1.09(1149)                & 1.06(612)                &&  0.85(439)             & 1.01(354)                        \\
\hline
\end{tabular}
\end{center}
\end{table*}

\subsubsection{\textit{Swift} and \textit{NuSTAR} spectral fits}  \label{sec:swnuspec}
We have also investigated IGR1509 by the spectral fitting of the contemporaneous \textit{Swift}-XRT and \textit{NuSTAR}-FPMA/FPMB observations of the epoch 2018 in the 0.3-10.0 keV and 3-78 keV energy bands, respectively, and are shown in Fig. \ref{fig:szswnuspec}. The iron line complex region is also seen in both the \textit{Swift} and 
\textit{NuSTAR} spectra. Similar to the approach as described for the \textit{Suzaku} spectral fitting (see Section \ref{sec:szspec}), we first investigated the \textit{Swift} and \textit{NuSTAR} spectra in the 0.3-78.0 keV energy band using model A which provides a better fit to the spectra. A large EW of the Fe K$\alpha$ emission was also observed in the \textit{Swift} and \textit{NuSTAR} spectra which further provides a hint of occurrence of X-ray reflection in the system. Therefore, to take into account the occurrence of the X-ray reflection and iron line complex region, we have used a \texttt{reflect} model as model B. With model B, we have found a better fit, but again with the unphysical value of $R_{refl}$ $>$ 1. Thus, we fixed this parameter to unity, allowing cos$(i)$ to vary as model C which provides a more satisfactory fit with $\chi^2_\nu$ of 0.95. In this way, we have adopted model C as a best-fit model for the \textit{Swift} and \textit{NuSTAR} spectra of the epoch 2018. The unabsorbed X-ray flux in the 0.3-78.0 keV energy band was also calculated for all these models. The best-fit parameters as obtained for each model are given in Table \ref{tab:xrayspec}.

In order to see the role of absorbers at pulse maximum (0.8-1.0) and pulse minimum (0.5-0.7) phases, the \textit{Suzaku}-XIS and contemporaneous \textit{Swift} and \textit{NuSTAR} spectra were also extracted at these phases. The spectral fitting was performed by using the best-fit model C as described above for average spectra, keeping fixed the Galactic absorption component, plasma temperatures and abundance values of the \texttt{mkcflow}, and the parameters of the blackbody components at the values obtained from the average spectral fitting. We have also used the same fixed parameters which were adopted for their average spectral fitting. The free parameters were pwab components ($N_{H,max}$ and $\beta$), the plasma normalizations ($n_{mkcflow}$), and the normalization of the Gaussian component ($n_{g6.4}$). Unabsorbed X-ray fluxes were also derived in the 0.3-10.0 keV and 0.3-78.0 keV energy bands for the \textit{Suzaku}-XIS and contemporaneous \textit{Swift} and \textit{NuSTAR} spectra, respectively. Resulting spectral parameters are summarized in Table \ref{tab:szswnuspinprs}, where the error bars are quoted with a 90 per cent confidence limit for a single variable parameter.


\section{Discussion} \label{sec:diss}
Based on the wealth of optical and X-ray observations, we explored the detailed optical and X-ray properties of an IP IGR1509. We have derived three prominent X-ray and optical frequencies at $\omega$, $2\omega$, and $3\omega$ which are well-consistent with the frequencies obtained from the previous X-ray and optical observations \citep[see][]{Pretorius09, Butters09, Potter12, Bernardini12}. Using the longer spanned \textit {TESS} light curves, we refined the spin period more precisely as 809.49584$\pm$0.00075 s. Additionally, the significant $\omega$$-$$\Omega$ frequency is present in the optical band and also a marginal detection in X-rays. This is for the first time, we have detected an unambiguous $\omega$$-$$\Omega$ frequency in the study of IGR1509. The presence of $\omega-\Omega$ seems intrinsic because if $\omega-\Omega$ would have been the orbital modulation of $\omega$, then $\omega+\Omega$ should be present in the power spectrum with almost same power as $\omega-\Omega$, which is contrary to what we observed. Hence, the $\omega-\Omega$ modulation seems to be originating from the accretion through the stream. Additionally, $\Omega$ and 3 ($\omega$$-2$$\Omega$) frequencies are also present in the \textit {TESS} optical power spectrum, unlike X-rays. The obscurations of the WD by the material rotating in the binary frame or an eclipse of the hotspot by an optically thick disc may be responsible for the orbital modulation \citep{Warner86}. However, non-detection of the orbital modulation in X-rays is not surprising, since orbital modulation in IPs are largely confined to soft energy bands. Moreover, four other side-band frequencies $\omega$$-$$3$$\Omega$, $\omega$$+$$3$$\Omega$, $2\omega$$-$$3$$\Omega$, and $2\omega$$+$$3$$\Omega$ located at both sides of frequencies $\omega$ and $2\omega$ are present in the \textit{NuSTAR} and \textit{Swift} power spectrum. The origin of these frequencies could be due to the amplitude modulation of the frequencies $\omega$ and $2\omega$ at thrice of the orbital frequency \citep[see][]{Warner86}. Unfortunately, their spacing does not correspond to one-third of the orbital period as well as the peak power of these frequencies is not always nearly the same, which provides a hint that these might be the spacecraft orbital aliases. All the observed X-ray and optical periodicities undoubtedly confirm the IP nature of IGR1509. Also, the presence of strong spin modulation in X-ray and optical bands indicates that the IGR1509 unveils the disc-fed dominance accretion. However, the detection of an additional beat frequency indicates that part of the accreting material also flows directly toward the WD along the magnetic field lines \citep[for detail see][]{Wynn92, Ferrario99}. 

X-ray modulation at the spin frequency is the deﬁnitive characteristic of the IPs, which can arise due to the two mechanisms: one is photoelectric absorption or electron scattering in the infalling material \citep{Rosen88, Rosen92, Kim95} and the second one is self-occultation of emission regions by the WD \citep{King84}. The observed X-ray spin pulse profiles in IGR1509 are found to be energy dependent, where strong modulations are observed in the soft X-ray energy bands. Also, the hardness ratio curve displays a strong modulation which are anti-correlated with respect to the intensity modulation. Therefore, the first possibility, \textit {i.e.,} the photoelectric absorption seems to be most feasible for the X-ray spin modulation in IGR1509, similar to the majority of IPs validating the ``accretion curtain" model \citep[see][for details]{Norton89}. In this model, the spin modulations are minimum when the accretion curtain points towards the observer where the absorption effect is maximum, and vice-versa. Thus, the observed strong soft energy ($<$ 10 keV) rotational modulations and hardening at the minimum intensity can be attributed to the accretion curtain scenario and is generally explained with variable complex absorbers. On the other hand, either tall shocks or reflection is generally thought to be responsible for strong hard X-ray spin modulations ($>$10 keV) in the IPs depending on phased or anti-phased spin light curve variations with respect to a low energy pulsation \citep{Mukai99, Demartino01}. However, if the spin modulations are of the order of 10\% or small, the complex absorbers may be responsible to produce a spin modulation of such amplitudes via Compton scattering \citep{Rosen92}. The amplitudes of spin modulations of IGR1509 are close to this order or less at hard energies, so this possibility seems to be the most feasible rather than the tall shocks or reflection. Considering also no significant hard energy spin modulation, we infer that the height of the X-ray emitting region may be negligible or closer to the WD surface. Moreover, the X-ray light curves are double-humped with a pronounced dip near phases $\sim$0.5-0.7. Based on the double-humped feature one can assume IGR1509 to be a two-pole accretor, but our analysis reveals that the dip between these phases is due to the photoelectric absorption in the intervening accretion stream. This is also evident from the hardness ratio curve, where the hardness ratio curve shows a substantial increase near the pronounced dip. 
 The amount of absorption is also found to be large 
  at this prominent dip phase of 0.5-0.7 (see Table \ref{tab:szswnuspinprs}). This, in turn, minimizes the possibility of a second accretion region or accretion from the second pole. No sign of cyclotron emission from the negative pole or detection of positive-only circular polarization reported by \cite{Potter12} also suggests one-pole accretion geometry in IGR1509. The observed single-peaked spin modulations in optical domain can be interpreted with the standard accretion-curtain model. In this model, optical emission originates from the accretion curtains between the inner disc and the WD. If they are optically thick, their varying aspect produces modulations during WD rotation. In this case, a single-peaked pulse with a maximum is observed when the upper pole points away from the observer \citep{Hellier91, Hellier95,  Kim96}.

X-ray spectra obtained from the \textit{Suzaku} and the contemporaneous \textit{Swift} and \textit{NuSTAR} observations in the 0.3-50.0 keV and 0.3-78.0 keV energy bands are well-modeled by a thick absorber with average hydrogen column densities of $>$4.4$\times$10$^{23}$ cm$^{-2}$ and $\sim$11$\times$$10^{23}$ cm$^{-2}$ with power-law indexes of -0.70 and -0.72 for covering fraction, a multi-temperature cooling flow model at maximum temperatures of $\sim$30 keV and $\sim$27 keV, along with blackbody temperatures of $\sim$89 eV and $\sim$81 eV, respectively. The hydrogen column density is found to be slightly higher than that previously reported by \cite{Bernardini12}. On the other hand, the maximum temperature appears to be close to their derived high temperature component. Similar to IGR1509, a high blackbody temperature was also observed in some other IPs \citep[see][etc.]{Haberl02,  Martino04, Evans07,  Anzolin08, Joshi19}, which are generally referred to as the soft IPs. The observed blackbody temperature 
 more than 80 eV places IGR1509 in the class of these soft-IPs, which can be explained with the heated region near the accretion footprints which is not hidden by the accretion curtains depending upon the system inclination and the magnetic colatitude \citep[see][]{Evans07}. The strong circular polarization reported by \cite{Potter12} also supports the presence of the soft X-ray emission in IGR1509. A similar feature was observed by \cite{Evans04} for the IPs PQ Gem, V405 Aur, and V2400 Oph, which reveals a strong polarization and possesses a soft blackbody emission component. The lack of blackbody emission during previous \textit {XMM-Newton} observations of IGR1509 could be attributed to a more dispersed accretion region, which can reduce the local heating rate and cause a decrease in the temperature of the soft X-ray component. We have also derived unabsorbed soft ($F_s$) X-ray fluxes of 2.1$_{-0.1}^{+0.1}$$\times$$10^{-11}$ erg cm$^{-2}$ s$^{-1}$ and 4.0$_{-0.6}^{+0.5}$$\times$$10^{-11}$ erg cm$^{-2}$ s$^{-1}$ from the blackbody model
 in the 0.3-50.0 keV and 0.3-78.0 keV energy bands, respectively, and hard ($F_h$) X-ray fluxes of 4.43$_{-0.03}^{+0.03}$$\times$$10^{-11}$ erg cm$^{-2}$ s$^{-1}$ and 5.70$_{-0.04}^{+0.04}$$\times$$10^{-11}$ erg cm$^{-2}$ s$^{-1}$ from the \texttt{mkcflow} model in the 0.3-50.0 keV and 0.3-78.0 keV energy bands, respectively. The softness ratio, $F_s$/$4F_h$, was then calculated as $\sim$0.12 and $\sim$0.17 for \textit{Suzaku} and the contemporaneous \textit{Swift} and \textit{NuSTAR} observations, respectively, which seems close to the softness ratios observed for some other soft-IPs \citep[see][]{Evans07}. We have also determined the size of the accretion footprint from the soft X-ray flux. The unabsorbed soft X-ray fluxes of 2.1$\times$$10^{-11}$ erg cm$^{-2}$ s$^{-1}$ and 4$\times$$10^{-11}$ erg cm$^{-2}$ s$^{-1}$ and the temperatures of 89 eV and 81 eV implies an emitting area of $\sim$1$\times$$10^{14}$ cm$^{2}$ and $\sim$1.3$\times$$10^{14}$ cm$^{2}$ for the \textit{Suzaku} and the contemporaneous \textit{Swift} and \textit{NuSTAR} observations, respectively. The WD mass of 0.73$\pm$0.06 $M_\odot$ \citep{Shaw20} thus implies the observed blackbody emitting area covers an order of $\sim$10$^{-5}$ of the WD surface of IGR1509. This area is consistent with other estimates for the accretion area in IPs \citep{Hellier97_3}. The phase-dependent analysis reveals that the absorption component is anti-correlated with the X-ray flux or plasma normalization, that is, the absorption component is found to be large at the pulse minimum phase and small at the pulse maximum phase (see Table \ref{tab:szswnuspinprs}). Such variations are compatible with the widely accepted classical curtain scenario \citep{Rosen88}. The phase-dependent variability is also seen in the fluorescent emission line components $n_{g6.4}$ and $EW_{g6.4}$. The observed values of $n_{g6.4}$ and $EW_{g6.4}$ are lower during the epoch 2011 of the \textit{Suzaku} observations than the epoch 2018 of the \textit{Swift} and \textit{NuSTAR} observations. This could be due to a different viewing angle onto the accretion column or due to the variable abundance of the ambient medium during different epoch of observations \citep[see][]{Schwope20}. 


\section{Conclusions} \label{sec:conc}
To conclude, we find the following characteristics of IGR1509, which are well explained by present optical and X-ray observations $-$
\begin{enumerate}
\item A more precise spin period of 809.49584$\pm$0.00075 s and orbital period of 5.87213$\pm$0.00014 h is derived using the long-baseline high-cadence optical photometric \textit {TESS} observations which are well consistent with the previously reported values. 
\item For the first time, an unambiguous beat period of 841.67376$\pm$0.00082 s is detected in IGR1509. 
\item Although, the detection of strong X-ray and optical frequencies $\omega$, $2\omega$, and $3\omega$ suggests that IGR1509 might be accreting predominantly via disc, however, the detection of an additional beat frequency in the present data indicates that part of the accreting material also flows directly toward the WD along the magnetic field lines.
\item The photoelectric absorption appears responsible for the soft X-ray ($<$10 keV) modulation. However, the complex absorbers may responsible to produce low amplitude spin modulations via Compton scattering in the  hard ($>$10 keV) energy band and indicates that the height of the X-ray emitting region may be negligible or close to the WD surface.
\item The observed double-humped X-ray profiles with a pronounced dip is indicative of the photoelectric absorption in the intervening accretion stream.
\item A multi-temperature plasma component, intrinsic absorption, reflection, and a soft component are found to be viable to explain the average X-ray spectra. The soft X-ray blackbody temperature is found to be in the range of 80$-$90 eV which places this system in the class of soft IPs. The estimated blackbody emitting area covers $\sim$10$^{-5}$ of the WD surface of IGR1509.
\end{enumerate}



\section*{Acknowledgments}
 We are grateful to the anonymous referee for providing helpful suggestions. This paper includes data collected with the \textit{TESS} mission, obtained from the MAST data archive at the Space Telescope Science Institute (STScI). Funding for the \textit{TESS} mission is provided by the NASA Explorer Program. We are grateful to all \textit{Suzaku} team members for their efforts in the production and the maintenance of the instruments and software, spacecraft operation, and calibrations. This research also use of data obtained with the \textit{NuSTAR} mission, a project led by the California Institute of Technology (Caltech), managed by the Jet Propulsion Laboratory (JPL) and funded by NASA. This research has made use of the XRT Data Analysis Software (XRTDAS) developed under the responsibility of the ASI Science Data Center (ASDC), Italy.  We acknowledge with thanks the variable star observations from the AAVSO International Database contributed by observers worldwide and used in this research.


\section*{DATA AVAILABILITY}
The \textit {Suzaku}, \textit {NuSTAR}, and \textit {Swift} data used for analysis in this article are publicly available in NASA’s High Energy Astrophysics Science Archive Research Center (HEASARC) archive (\url{https://heasarc.gsfc.nasa.gov/docs/archive.html}). The \textit{TESS} data sets are publicly available in the \textit{TESS} data archive at \url{https://archive.stsci.edu/missions-and-data/tess}. The AAVSO and ASAS-SN data sets are available at \url{https://www.aavso.org/data-download} and \url{https://asas-sn.osu.edu/variables}, respectively.

\bibliographystyle{mnras}
\bibliography{ref}
\label{lastpage}
\end{document}